\documentclass[sigplan, 10pt]{acmart}

\setcopyright{acmlicensed}
\copyrightyear{2018}
\acmYear{2018}
\acmDOI{XXXXXXX.XXXXXXX}
\acmConference[Conference acronym 'XX]{Make sure to enter the correct
  conference title from your rights confirmation email}{June 03--05,
  2018}{Woodstock, NY}
\acmISBN{978-1-4503-XXXX-X/2018/06}

\renewcommand\footnotetextcopyrightpermission[1]{}

\renewcommand\footnotetextcopyrightpermission[1]{}
\usepackage[utf8]{inputenc}
\usepackage{microtype}
\usepackage{xcolor}
\usepackage{tikz}
\usepackage{amsmath}
\usepackage{enumitem}
\usepackage{pifont}
\usepackage{xspace}
\usepackage{hyphenat}
\usepackage{stringstrings}
\usepackage{booktabs}
\usepackage{multirow}
\usepackage{array}
\usepackage{caption}
\usepackage{subcaption}
\usepackage{adjustbox}
\usepackage{url}

\AtBeginDocument{%
  \setlength{\abovedisplayskip}{3pt plus 1pt minus 1pt}%
  \setlength{\belowdisplayskip}{3pt plus 1pt minus 1pt}%
  \setlength{\abovedisplayshortskip}{2pt plus 1pt}%
  \setlength{\belowdisplayshortskip}{2pt plus 1pt}%
}

\definecolor{MyGreen}{RGB}{1,150,32}
\definecolor{MyRed}{RGB}{150,1,32}

\newcommand{\sys}{\textsc{Tangram}\xspace}
\newcommand{\msi}{model\hyp slice\slash island\xspace}

\newcommand{\cmark}{\textcolor{MyGreen}{\ding{51}}}

\newcommand{\xmark}{\textcolor{MyRed}{\ding{55}}}

\newcommand{\ie}{i.e.,\ }
\newcommand{\eg}{e.g.,\ }
\newcommand{\unit}[1]{\mbox{\hspace{2pt}#1}\xspace}
\newcommand{\F}{\mbox{Fig.\hspace{0.25em}}}
\newcommand{\T}{\mbox{Tab.\hspace{0.25em}}}
\newcommand{\code}[1]{\texttt{\small{}#1}\xspace}

\newcommand*\myc[1]{%
\scalebox{0.78}{\begin{tikzpicture}[baseline=-3pt]
  \protect\node[draw,circle,inner sep=0.5pt, fill=black] {\textcolor{white}{\textsf{\textbf{#1}}}};
\end{tikzpicture}}}

\newcommand{\tinyskip}{\vspace{2pt}}
\newcommand{\mypar}[1]{\tinyskip\noindent\textbf{#1.}\xspace}
\newcommand{\myparr}[1]{\tinyskip\noindent\textbf{#1}\xspace}

\newenvironment{myitemize}{%
\begin{itemize}[leftmargin=1em, itemsep=.1em, parsep=.1em, topsep=.1em,
    partopsep=.1em]}
{\end{itemize}}

\newenvironment{myenumerate}{%
\begin{enumerate}[leftmargin=1.50em, itemsep=0em, parsep=0em, topsep=.1em,
    partopsep=.1em]}
{\end{enumerate}}

\begin{document}

\date{}

\title{\sys{}: Hiding GPU Heterogeneity for Efficient\\ LLM Parallelization}

\author{
  \rm
    Yanda Tao$^{\dagger}$ \quad
    Pedro F. Silvestre$^{\dagger}$ \quad
    Marcel Wagenl{\"a}nder$^{\dagger}$ \quad
    Peter Pietzuch$^{\dagger}$ \\[4pt]
  {\small
    $^{\dagger}$Imperial College London \quad
  }
}

\renewcommand{\shortauthors}{Tao et al.}


\begin{abstract}
  The scale of LLM training jobs requires parallelization planning over large GPU clusters. Due to different GPU types and interconnects added over time, these GPU clusters are increasingly heterogeneous. Automatic LLM parallelizers can search for parallelization plans but face an exploding search space with heterogeneous GPUs. To make search tractable in heterogeneous GPU clusters, parallelizers often omit types of parallelism (\eg expert parallelism) or memory-saving techniques (\eg ZeRO), which results in worse plans.

    We describe \sys, a system that enables the use of existing heterogeneity-unaware LLM parallelizers in heterogeneous GPU clusters by decoupling parallelization planning from GPU heterogeneity. For this, \sys exploits two insights: (1)~since bulk purchases result in sets of GPUs with similar compute, memory, and connectivity, \sys can expose such homogeneous \emph{GPU islands} to existing parallelizers; and (2)~parallelizers commonly first partition models and then parallelize partitions. \sys{} can compose such \emph{model slices}, assigned to GPU islands, into work-balanced pipelines for high throughput. \sys integrates with existing parallelizers through a narrow API, which relies on the enumeration of model-slice/island pairs. \sys achieves up to $2.3\times$ higher training throughput than current heterogeneous parallelizers (Metis and Sailor) and scales to large GPU clusters by pruning enumerated plans.
\end{abstract}


\settopmatter{printfolios=true,printacmref=false}

\pagestyle{plain}
\maketitle
\thispagestyle{plain}


\section{Introduction}
\label{sec:intro}

\begin{figure}[t]
  \centering
  \includegraphics[width=0.98\columnwidth]{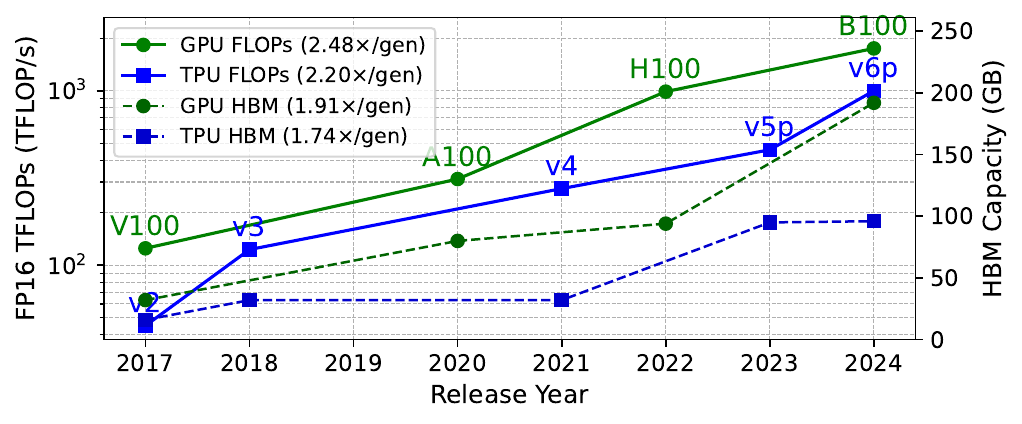}
  \caption{GPU compute and memory improvements by year. \textnormal{(Previous GPU generations remain competitive in terms of floating point operations per second~(FLOPs) and high-bandwidth memory~(HBM) capacity.)}}
  \label{fig:xpus}
\end{figure}

Large-language models~(LLMs) follow empirical scaling laws: with more compute and data used during training, LLMs become more capable~\cite{hoffmann2022training}. Organizations therefore regularly expand their GPU clusters by upgrading to the latest GPU hardware~\cite{aiindex2025}. \F\ref{fig:xpus} shows that each GPU generation improves compute and memory capacities by a factor of $\sim$$2\times$. Since previous GPU generations still offer substantial performance, organizations keep them to amortize capital expenditure~\cite{semianalysis2025gpulifetimes}, which results in \emph{heterogeneous} GPU clusters~\cite{stojkovic2025rearchitecting}.

LLMs have grown to trillions of parameters~\cite{fedus2022switch, liu2024deepseek, deepseekv4}, and their training must use the aggregate compute and memory resources of large, potentially heterogeneous, GPU clusters. For this, LLM training relies on \emph{multi-dimensional parallelism}: data~\cite{li2020pytorch} and context parallelisms~\cite{liu2023ring} partition the training data; and tensor~\cite{shoeybi2019megatron}, pipeline~\cite{huang2019gpipe, narayanan2019pipedream}, and expert parallelisms~\cite{fedus2022switch, lepikhin2020gshard} partition the model. In addition, memory-saving techniques~\cite{micikevicius2017mixed, rajbhandari2020zero, rhu2016vdnn, chen2016training, pytorchfsdp} trade off compute, accuracy, or bandwidth for lower GPU memory usage. Modern LLM training systems~\cite{shoeybi2019megatron, rasley2020deepspeed, openxla2025pjrt} combine these strategies into \emph{parallelization plans}~\cite{llama3}. Choosing an effective plan is crucial for good training performance, and doing so manually is hard with many parallelization strategies, complex model architectures and heterogeneous GPUs~\cite{strati2025sailor}.

Automatic \emph{LLM parallelizers}~\cite{fan2021dapple, zheng2022alpa, miao2022galvatron, liu2024aceso, zhu2025mist, li2025hypertron} use profiling, cost modeling, and search algorithms to explore possible parallelization plans, and return plans that result in high training throughput. Most existing parallelizers assume homogeneous GPUs, which makes the parallelization problem symmetric across GPUs: training throughput depends only on the plan and not the GPU placement, \ie the mapping of model and data partitions to specific GPUs. Hence, such \emph{homogeneous} parallelizers can exploit efficient search strategies for finding good parallelization plans, \eg considering intra- and inter-operator plans hierarchically~\cite{miao2022galvatron, zheng2022alpa, zhu2025mist}, or using beam search with iterative bottleneck alleviation~\cite{liu2024aceso}. In heterogeneous GPU clusters, however, without accounting for different GPU types (see~\F\ref{fig:xpus}) or interconnect bandwidths~\cite{nvlink, pcie, ethernet, infiniband}, training throughput may become bottlenecked by overloaded slower GPUs or interconnects~\cite{strati2025sailor}.

In response, \emph{heterogeneous} LLM parallelizers~\cite{um2024metis, strati2025sailor, harp, hetauto, hexiscale} try to account for heterogeneous GPUs in their search, jointly optimizing parallelisms and GPU placement. This drastically increases the search space size by several orders of magnitude~(see~\F\ref{fig:search_space_incremental}): for each parallelism, the parallelizer must consider the mapping of model and data partitions to each heterogeneous GPU. To counter this search space explosion, heterogeneous LLM parallelizers either omit parallelisms and/or memory-saving techniques (\eg expert parallelism~\cite{fedus2022switch} or ZeRO~\cite{rajbhandari2020zero}), or aggressively prune potential plans. For example, Metis~\cite{um2024metis}, a recent heterogeneous LLM parallelizer, is outperformed by up to 2$\times$ in training throughput by a state-of-the-art homogeneous parallelizer, Aceso~\cite{liu2024aceso}, even in heterogeneous GPU clusters: while Aceso ignores heterogeneity, it supports per-operator activation recomputation~\cite{chen2016training}, which yields better plans than Metis. Aceso's smaller search space allows for such features without significantly increasing search time, which is infeasible with Metis's larger search space due to GPU heterogeneity.

Therefore, due to their smaller search spaces when considering parallelization plans, homogeneous LLM parallelizers consistently provide more comprehensive support for the latest parallelisms and memory-saving techniques compared to heterogeneous LLM parallelizers. For example, recent heterogeneous parallelizers such as Metis~\cite{um2024metis}, Sailor~\cite{strati2025sailor}, and Hexiscale~\cite{hexiscale} do not support context and expert parallelisms; AMP~\cite{li2022amp} and HetAuto~\cite{hetauto} do not consider activation recomputation or ZeRO~\cite{rajbhandari2020zero}. Simply adding support for such features would increase the size of their search spaces to a point at which search times would become intractable~(see~\S\ref{sec:het_challenges}). 

\begin{figure}[t]
  \centering
  \includegraphics[width=\columnwidth]{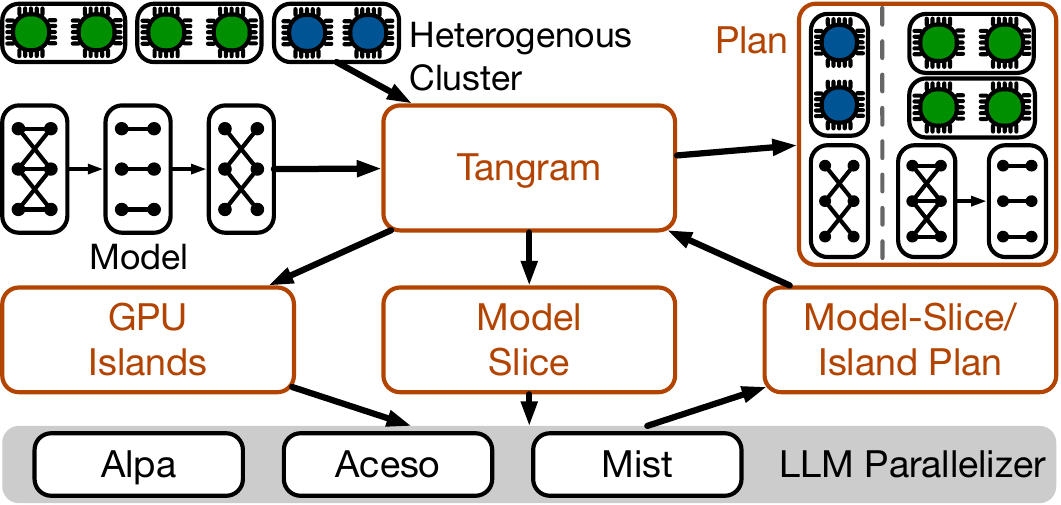}
  \caption{\sys{} interfacing with any LLM parallelizer}\label{fig:hetml_narrow_waist}
\end{figure}

To solve this challenge, our key insight is that parallelization can be decoupled from GPU heterogeneity: taking full advantage of their feature sets, homogeneous LLM parallelizers can generate \emph{partial} plans by parallelizing a slice of a model on a subset of homogeneous GPUs. These partial plans can be composed in a manner that accounts for the GPU heterogeneity in a cluster. Based on this idea of hiding GPU heterogeneity from LLM parallelizers, we describe \sys{}\footnote{Available at \url{https://github.com/}\emph{<anonymized>}}, a system that enables the use of existing homogeneous LLM parallelizers to produce high-throughput parallelization plans for heterogeneous GPU clusters (see \F\ref{fig:hetml_narrow_waist}). \sys constructs a heterogeneous plan out of enumerated partial plans from homogeneous LLM parallelizers. It then composes these partial plans through a pipeline that balances work among heterogeneous GPUs. In more detail, \sys makes the following technical contributions:

\mypar{(1)~GPU island abstraction~(\S\ref{sec:cluster_decomposition})} To utilize existing LLM parallelizers without breaking their assumptions about GPU homogeneity, \sys provides them with homogeneous GPUs through the abstraction of a homogeneous \emph{GPU island}. A GPU island is constructed by grouping GPU nodes based on their performance characteristics: \sys clusters GPUs according to intra-node metrics, such as compute and memory capacities, and inter-node metrics, such as network connectivity, to form GPU islands. Since the number of GPU islands impacts planning time, \sys keeps island counts low by greedily merging GPU islands whose differences in performance metrics can be tolerated, \eg inter-node interconnects with heterogeneous, but high bandwidth. This does not impact planning quality because intra-island GPU heterogeneity is still constrained to avoid bottlenecks.

\mypar{(2)~Generic LLM parallelizer interface~(\S\ref{sec:model_slicing})} To integrate with diverse homogeneous LLM parallelizers, \sys requires a generic interface. Parallelizers partition LLMs at different granularities (\eg layer vs.\ operator) and levels (\eg full computational graph vs.\ forward-layers only). \sys exploits this by providing a \emph{model slice} as the input to the parallelizer, which encapsulates a part of the model. Since model slices are similar to the full model with inputs, outputs, and layers, they are compatible with existing parallelizers. The parallelizer is thus given a model slice and GPU island, and it returns \emph{model-slice/island plans}, which are partial parallelization plans for that model slice, assuming the homogeneous GPU island.

\mypar{(3)~Pruning model-slice/island exploration~(\S\ref{sec:slice_plan_generation})} With large LLMs and GPU cluster sizes, the combinatorial number of pairs of model-slices and GPU islands becomes intractable. When \sys enumerates model-slice/island pairs to request partial plans from an LLM parallelizer, it applies pruning policies: it discards \emph{redundant pairs}, \eg model slices with the same number of transformer layers, and \emph{unbalanced pairs}, \ie those that would assign a large model slice to a resource-limited GPU island. In addition, \sys ignores \emph{infeasible pairs}, \eg due to GPU memory constraints, and \emph{incompatible partial plans}, which cannot be composed into a global plan. As we show in \S\ref{sec:eval:pruning}, these pruning policies reduce the time spent on plan generation by $26\times$.

\begin{figure*}[t]
  \centering
  \begin{subfigure}[b]{0.26\textwidth}
    \centering
    \includegraphics[width=\textwidth, keepaspectratio]{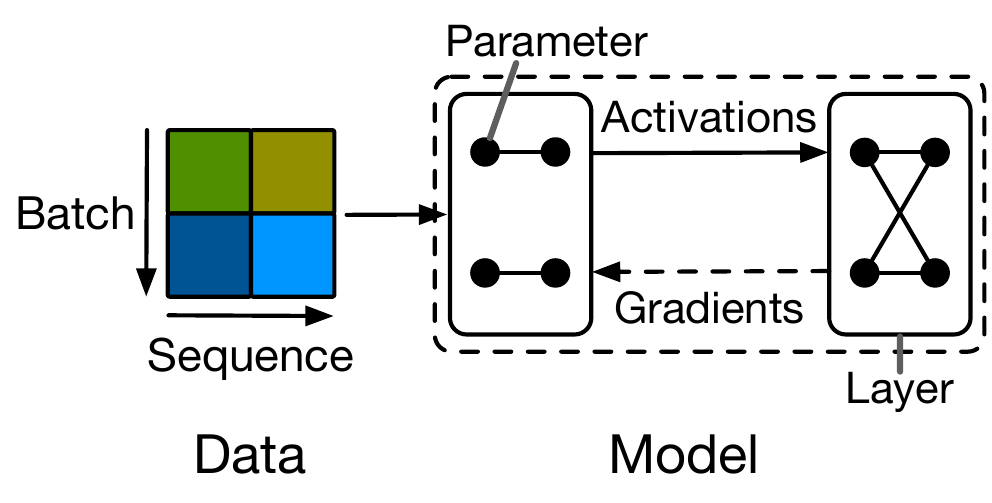}
    \caption{LLM training}
    \label{fig:tech:base}
  \end{subfigure}
  \hfill
  \begin{subfigure}[b]{0.16\textwidth}
    \centering
    \includegraphics[width=\textwidth, height=2cm, keepaspectratio]{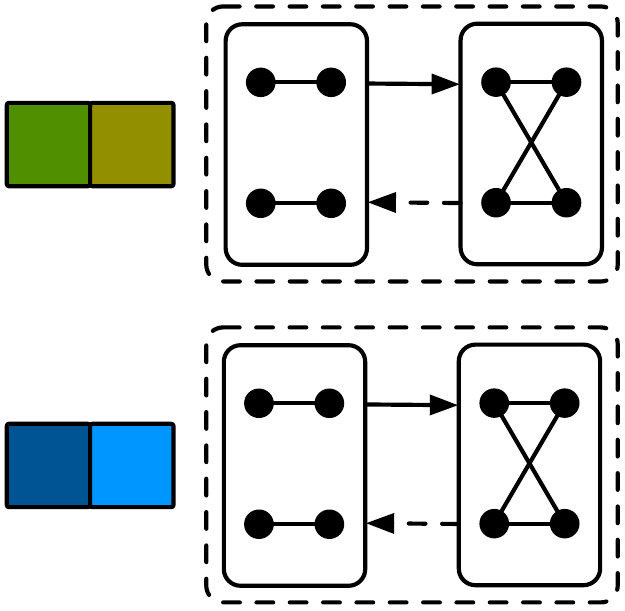}
    \caption{Data parallelism~(DP)}
    \label{fig:tech:dp}
  \end{subfigure}
  \hfill
  \begin{subfigure}[b]{0.16\textwidth}
    \centering
    \includegraphics[width=\textwidth, height=2cm, keepaspectratio]{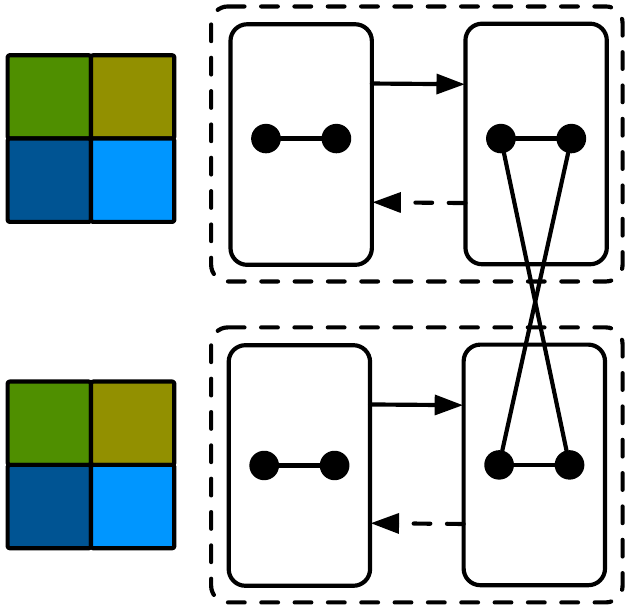}
    \caption{Tensor parallelism~(TP)}
    \label{fig:tech:tp}
  \end{subfigure}
  \hfill
  \begin{subfigure}[b]{0.16\textwidth}
    \centering
    \includegraphics[width=\textwidth, height=2cm, keepaspectratio]{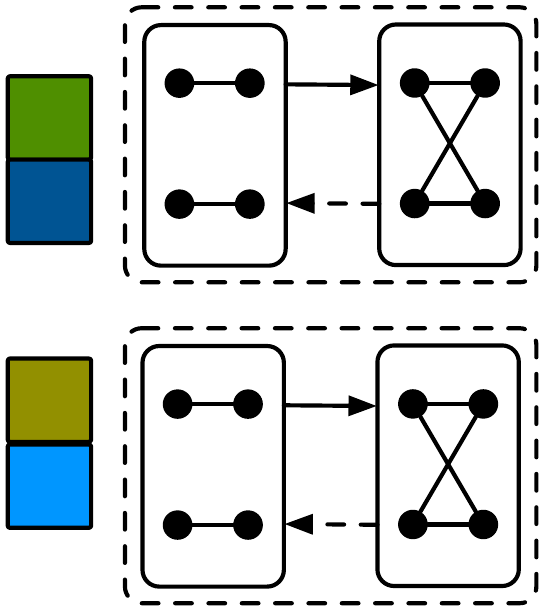}
    \caption{Context parallelism~(CP)}
    \label{fig:tech:cp}
  \end{subfigure}
  \hfill
  \begin{subfigure}[b]{0.16\textwidth}
    \centering
    \includegraphics[width=\textwidth, height=2cm, keepaspectratio]{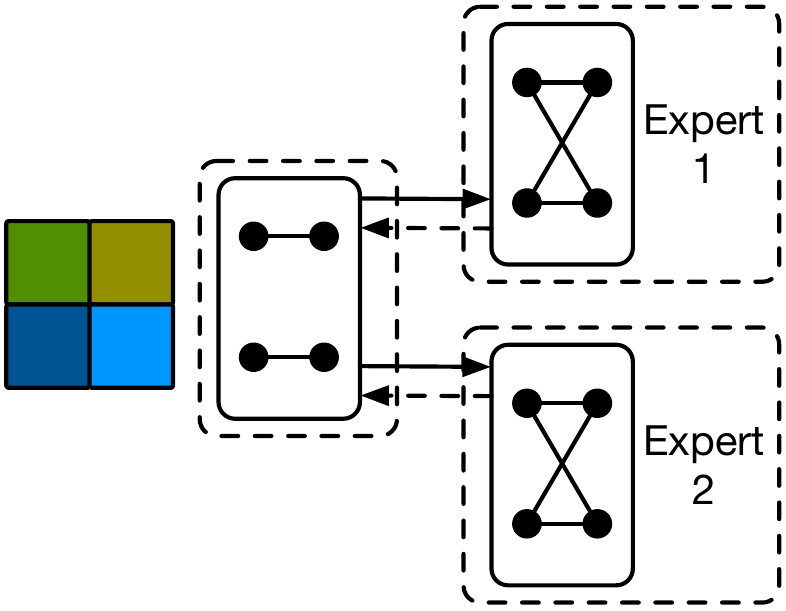}
    \caption{Expert parallelism~(EP)}
    \label{fig:tech:ep}
  \end{subfigure}

  \vspace{0.25cm}

  \begin{subfigure}[b]{0.16\textwidth}
    \centering
    \includegraphics[width=\textwidth, height=2.25cm, keepaspectratio]{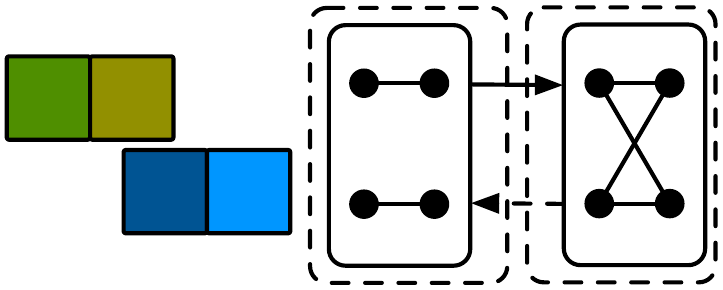}
    \caption{Pipeline parallelism~(PP)}
    \label{fig:tech:pp}
  \end{subfigure}
  \hfill
  \begin{subfigure}[b]{0.19\textwidth}
    \centering
    \includegraphics[width=\textwidth, height=1.5cm, keepaspectratio]{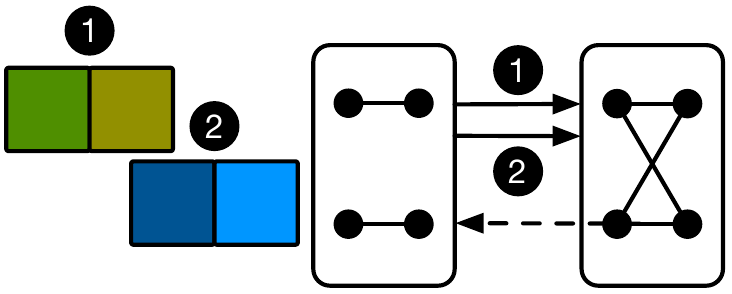}
    \caption{Gradient accumulation~(GA)}
    \label{fig:tech:grad_accum}
  \end{subfigure}
  \hfill
  \begin{subfigure}[b]{0.19\textwidth}
    \centering
    \includegraphics[width=\textwidth, height=1.5cm, keepaspectratio]{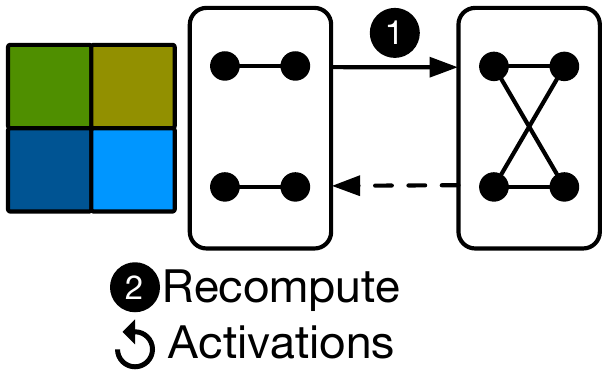}
    \caption{Activation recomputation~(RC)}
    \label{fig:tech:recomputation}
  \end{subfigure}
  \hfill
  \begin{subfigure}[b]{0.19\textwidth}
    \centering
    \includegraphics[width=\textwidth, height=1.5cm, keepaspectratio]{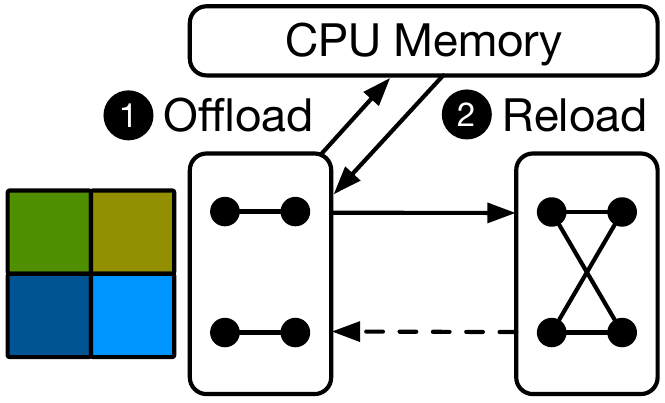}
    \caption{Memory offloading~(OF)}
    \label{fig:tech:offload}
  \end{subfigure}
  \hfill
  \begin{subfigure}[b]{0.19\textwidth}
    \centering
    \includegraphics[width=\textwidth, height=1.5cm, keepaspectratio]{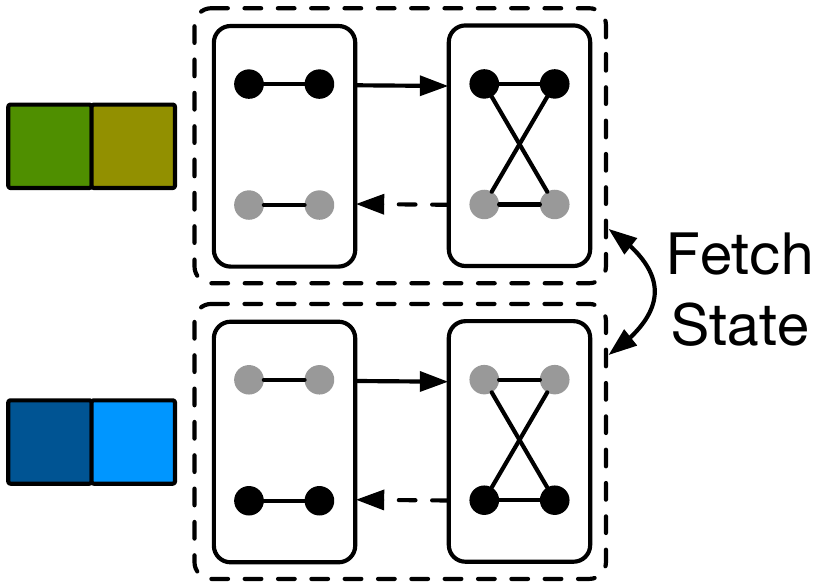}
    \caption{ZeRO/FSDP}
    \label{fig:tech:zero}
  \end{subfigure}

  \caption{Parallelisms and memory-saving techniques in distributed LLM training}
  \label{fig:techniques}
\end{figure*}

\mypar{(4)~Partial plan composition using pipeline stages~(\S\ref{sec:search_scalability})} \sys must compose the partial model-slice/island plans in a way that maximizes training throughput while accounting for the heterogeneity across different GPU islands. \sys achieves this by creating a \emph{pipeline} of balanced stages: each stage belongs to a model-slice/island plan, and all model slices in the pipeline must together compose the full model. Pipeline parallelism allows partial plans with arbitrary parallelisms to be composed, with low communication requirements between GPU islands. Since pipeline throughput is dominated by the slowest stage, \sys composes multiple feasible pipelines under different stage imbalances using a dynamic programming algorithm, and selects one that yields the highest training throughput.

\tinyskip

\noindent
We evaluate \sys (\S\ref{sec:evaluation}) in comparison to homogeneous (Alpa~\cite{zheng2022alpa}, Aceso~\cite{liu2024aceso}) and heterogeneous (Metis~\cite{um2024metis}, Sailor~\cite{strati2025sailor}) LLM parallelizers with different LLMs and GPU cluster configurations. On heterogeneous GPU clusters, \sys outperforms homogeneous and heterogeneous LLM parallelizers in training throughput by up to $2.1\times$ and $2.3\times$, respectively; on homogeneous GPU clusters, \sys matches the performance of homogeneous LLM parallelizers (Aceso) while outperforming heterogeneous ones (Metis and Sailor) by up to $3.1\times$. We also demonstrate \sys's compatibility with recent MoE parallelization strategies by integrating with Galvatron~\cite{miao2022galvatron}'s MoE runtime, LAER-MoE~\cite{laermoe}, on the Mixtral~\cite{mixtral} and Qwen3-MoE models~\cite{qwen3}.



\section{Parallelization in Heterogeneous GPU \mbox{Clusters}}

Next, we describe distributed LLM training and common parallelization techniques to scale to large GPU clusters~(\S\ref{sec:training}). After that, we introduce the challenges from GPU heterogeneity~(\S\ref{sec:het_challenges}), describe the design space of automatic LLM parallelizers~(\S\ref{sec:background:auto_llm_parallelizers}), and survey existing homogeneous and heterogeneous LLM parallelizers~(\S\ref{sec:background:homo_het_parallelizers}).

\subsection{Parallelization in distributed LLM training}
\label{sec:training}

To train an LLM, its parameters must be updated iteratively, \eg using stochastic gradient descent~(SGD)~\cite{sgd}. As shown in \F\ref{fig:tech:base}, each training iteration uses a batch of input tokens from a dataset as an input tensor and proceeds in three steps: (1) in a \emph{forward pass}, the batch is processed by each layer, producing \emph{activation} tensors passed to the next layer. The last layer produces a prediction, and a \emph{loss} value is calculated to quantify the prediction error based on the target output; (2) in a \emph{backward pass}, gradients are computed and propagated back through the LLM~\cite{rojas1996backpropagation}; and (3) in a \emph{parameter update} step, an optimizer, such as Adam~\cite{kingma2014adam}, takes the gradients and applies parameter updates to improve the model accuracy.

Modern LLMs~\cite{qwen3.5, deepseekv4} far exceed the compute and memory capacities of one GPU. To scale LLM training across many GPUs, parallelization strategies are used as part of sophisticated parallelization plans~\cite{llama3}. Existing strategies broadly fall into \emph{parallelisms} and \emph{memory-saving} techniques:

\myparr{Parallelisms} partition the model and data across GPUs for parallel execution: (i)~data parallelism~(DP in \F\ref{fig:tech:dp})~\cite{li2020pytorch} replicates model parameters to process data partitions in parallel, but must synchronize gradients at the end of each iteration; (ii)~tensor parallelism~(TP in \F\ref{fig:tech:tp})~\cite{shoeybi2019megatron} partitions model parameters, but must communicate intermediate outputs after each model layer; (iii)~context parallelism~(CP in \F\ref{fig:tech:cp})~\cite{liu2023ring} partitions input sequences to reduce memory requirements, but requires extra communication for attention computation; (iv)~expert parallelism~(EP in \F\ref{fig:tech:ep})~\cite{fedus2022switch} distributes experts in the mixture-of-experts~(MoE) layer across GPUs, but requires communication for routing tokens to activated experts and aggregating their outputs; and (v)~pipeline parallelism~(PP in \F\ref{fig:tech:pp})~\cite{huang2019gpipe, narayanan2019pipedream} partitions the model at operator/layer boundaries, forming pipeline stages that consume micro-batches of data.

\begin{table*}[t]
  \centering
  \caption{Survey of existing automatic LLM parallelizers}
  \label{tab:autopara_survey}
  \begin{adjustbox}{max width=\linewidth}
    \begin{tabular}{lllllcccccccccc}
      \toprule
      \multirow{2.4}{*}{\textbf{Category}} &
      \multirow{2.4}{*}{\textbf{System}}   &
      \multirow{2.4}{*}{\textbf{Year}}     &
      \multirow{2.4}{*}{\textbf{Runtime}}  &
      \multirow{2.4}{*}{\textbf{Search}}  &
      \multicolumn{5}{c|}{\textbf{Parallelism}} &
      \multicolumn{4}{c}{\textbf{Memory-saving}} &
      \multirow{2.4}{*}{\textbf{GPU Placement}}     \\
      \cmidrule(lr){6-10}\cmidrule(lr){10-14}
      &
      &
      &
      &
      &
      DP                                 & PP                                & TP                                & CP                                & EP                                &
      GA                         & RC                           & OF                           & ZeRO                           &   \\
      \midrule

      \multirow{5}{*}{\textbf{Homogeneous}}
      & Alpa~\cite{zheng2022alpa}        & 2022 & PJRT (XLA)      &
      DP + ILP &
      \cmark{}                         & \cmark{}                         & \cmark{}                         & \xmark{}                         & \cmark{}                         &
      \xmark{}                         & \xmark{}                         & \xmark{}                         & \xmark{}                         &
      \xmark{} \\
      & Galvatron~\cite{miao2022galvatron} & 2022 & Megatron (PT) &
      Decision-tree &
      \cmark{}                         & \cmark{}                         & \cmark{}                         & \cmark{}                         & \xmark{}                         &
      \xmark{}                         & \cmark{}                         & \xmark{}                         & \cmark{}                         &
      \xmark{} \\
      & Aceso~\cite{liu2024aceso}        & 2024 & Megatron (PT)   &
      Alleviation &
      \cmark{}                         & \cmark{}                         & \cmark{}                         & \xmark{}                         & \xmark{}                         &
      \xmark{}                         & \cmark{}                         & \xmark{}                         & \xmark{}                         &
      \xmark{} \\

      & Mist~\cite{zhu2025mist}          & 2025 & Custom (PT)              &
      MILP + enumeration &
      \cmark{}                         & \cmark{}                         & \cmark{}                         & \xmark{}                         & \xmark{}                         &
      \cmark{}                         & \cmark{}                         & \cmark{}                         & \cmark{}                         &
      \xmark{} \\
      & HyperTron~\cite{li2025hypertron} & 2025 & Megatron (PT)             &
      Annealing &
      \cmark{}                         & \cmark{}                         & \cmark{}                         & \cmark{}                         & \cmark{}                         &
      \xmark{}                         & \xmark{}                         & \xmark{}                         & \xmark{}                         &
      \xmark{} \\

      \midrule

      \multirow{6}{*}{\textbf{Heterogeneous}}

      & AMP~\cite{li2022amp}             & 2022 & DeepSpeed (PT)              &
      DP &
      \cmark{}                         & \cmark{}                         & \cmark{}                         & \xmark{}                         & \xmark{}                         &
      \xmark{}                         & \xmark{}                         & \xmark{}                         & \xmark{}                         &
      PP (No mem. est.) \\
      & Metis~\cite{um2024metis}         & 2024 & PJRT (XLA)              &
      DFS &
      \cmark{}                         & \cmark{}                         & \cmark{}                         & \xmark{}                         & \xmark{}                         &
      \xmark{}                         & \xmark{}                         & \xmark{}                         & \xmark{}                         &
      DP \& PP \\
      & Sailor~\cite{strati2025sailor}   & 2025 & DeepSpeed (PT)              &
      DP &
      \cmark{}                         & \cmark{}                         & \cmark{}                         & \xmark{}                         & \xmark{}                         &
      \xmark{}                         & \xmark{}                         & \xmark{}                         & \xmark{}                         &
      DP$^\ast$ \& PP \\
      & HARP~\cite{harp}   & 2026 & PJRT (XLA)              &
      DP + ILP &
      \cmark{}                         & \cmark{}                         & \cmark{}                         & \xmark{}                         & \cmark{}                         &
      \xmark{}                         & \xmark{}                         & \xmark{}                         & \xmark{}                         &
      PP \\
      & HetAuto~\cite{hetauto}   & 2026 & Megatron (PT)              &
      MCTS &
      \cmark{}                         & \cmark{}                         & \cmark{}                         & \cmark{}                         & \xmark{}                         &
      \xmark{}                         & \xmark{}                         & \xmark{}                         & \xmark{}                         &
      PP \\
      & HexiScale~\cite{hexiscale}   & 2026 & Megatron (PT)              &
      Hierarchical graph part. &
      \cmark{}                         & \cmark{}                         & \cmark{}                         & \xmark{}                         & \xmark{}                         &
      \xmark{}                         & \xmark{}                         & \xmark{}                         & \xmark{}                         &
      DP \& PP \\


      \bottomrule
    \end{tabular}
  \end{adjustbox}
\end{table*}

Since parallelisms partition and replicate different parts of the training state, they are combined as \emph{multi-dimensional parallelism}~(MDP) for the highest training throughput. The choice of an MDP configuration depends on the training scenario. For example, EP applies only to MoE LLMs~\cite{fedus2022switch,qwen3.5}; long-sequence training is more memory-intensive and typically requires CP~\cite{llama3} to accommodate its memory usage.

\myparr{Memory-saving} techniques reduce memory pressure during LLM training in exchange for extra overhead: (i)~gradient accumulation~(GA in \F\ref{fig:tech:grad_accum}) splits each batch into micro-batches and processes them sequentially, while accumulating gradients, which increases the effective batch size; (ii)~activation recomputation~(RC in \F\ref{fig:tech:recomputation})~\cite{chen2016training} frees memory by discarding forward pass activations and recomputing them during the backward pass; (iii)~activation offloading~(OF in \F\ref{fig:tech:offload})~\cite{ren2021zero} moves activations to CPU host memory between forward and backward passes; and (iv)~ZeRO-redundancy optimization~\cite{rajbhandari2020zero}~(in \F\ref{fig:tech:zero}), or FSDP for ZeRO-3~\cite{pytorchfsdp}, partitions optimizer states, gradients, and parameters across data-parallel GPUs, which avoids the peak memory usage of full replication. Memory-saving techniques are applied in combination with MDP~\cite{llama3} to balance trade-offs and achieve higher throughput.

\subsection{Challenges with heterogeneous GPU clusters}
\label{sec:het_challenges}

As GPU clusters grow, they also become more heterogeneous over time due to bulk purchases of the latest GPUs~\cite{semianalysis2025gpulifetimes}. In production GPU clusters, hardware heterogeneity arises not only from different GPU compute capabilities (\eg Nvidia Blackwell~\cite{blackwell} vs.\ Hopper~\cite{h100} GPUs) and memory capacities (\eg Nvidia H100 GPUs, SXM 80\unit{GB} vs.\ NVL 94\unit{GB})~\cite{MLaaS, bootseer}, but also from different interconnects~\cite{sai2024kalos} (\eg NVLink~\cite{nvlink} vs.\ PCIe~\cite{pcie} and Ethernet~\cite{ethernet} vs.\ Infiniband~\cite{infiniband}).

In heterogeneous GPU clusters, the choice of parallelization strategy and the assignment of data/model partitions to GPUs (GPU placement) become coupled. For example, when the interconnect bandwidth between two GPUs is low, common training recipes recommend against using TP across these GPUs, because TP requires substantial communication~\cite{zheng2022alpa, megatronbridge}. Therefore, a parallelization plan must consider GPU placement when selecting the parallelisms to use.

In addition, with GPU heterogeneity, parallelisms can no longer rely on uniform partitions~\cite{rasley2020deepspeed} and instead must use \emph{uneven partitioning}, where partitions account for heterogeneous GPU resources. For example, under PP, more resource-rich GPUs must be assigned more LLM layers for their pipeline stages, as long as memory constraints are met~\cite{um2024metis}; otherwise, GPUs would become idle while waiting for less capable GPUs during collective communication.

\begin{figure}[t]
  \centering
  \includegraphics[width=1.\columnwidth]{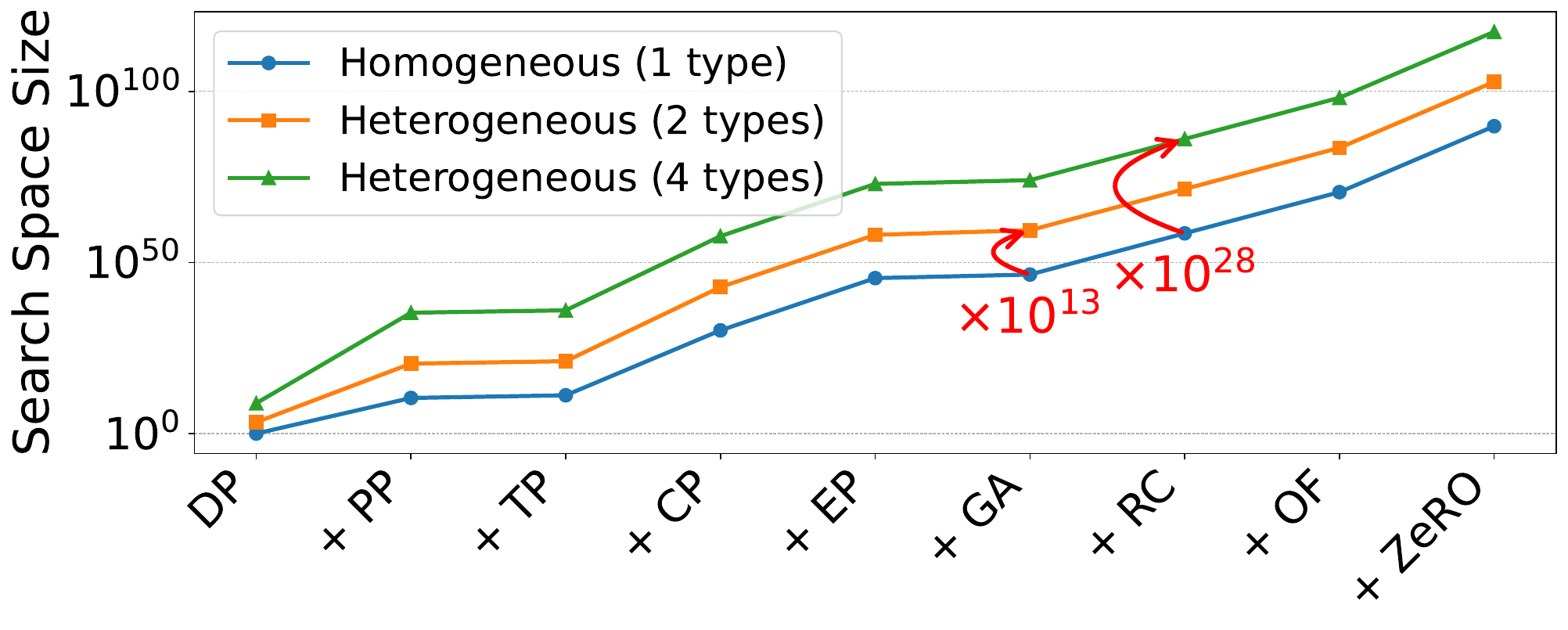}
  \caption{Growth of parallelization plans with heterogeneous GPUs \textnormal{(Assuming a GPT-3-13B model, each technique increases the configuration space by orders of magnitude, with heterogeneous GPU placement compounding this.)}}\label{fig:search_space_incremental}
\end{figure}

In \F\ref{fig:search_space_incremental}, we model the number of possible MDP plans for a 40-layer LLM on a 512-GPU cluster with homogeneous and heterogeneous GPUs. MDP configurations are obtained as tuples of parallelism degrees, which cover the entire cluster (\eg 8-way DP and 64-way PP). GA is picked globally, while ZeRO is chosen per-stage, and RC and OF are chosen per-layer. When placing on heterogeneous GPUs, we consider two options: each GPU type can process a different portion of a mini-batch (\ie uneven data partitioning), and each pipeline stage can be assigned to a different GPU type.

As \F\ref{fig:search_space_incremental} shows, the number of configurations increases exponentially with more parallelisms and memory-saving techniques. When the GPU cluster includes two heterogeneous GPU types, the search space size increases further by $10^{13}$. This increase is dominated by heterogeneous GPU placement across PP stages, which accounts for more than $10^9$ possibilities. If uneven PP stages are modeled, the increase becomes even more significant: for an LLM with $L$~layers and each PP degree~$S$, there can be $\binom{L-1}{S-1}$ configurations. The size of the configuration space makes it challenging for users to find a good solution, forcing them to rely on imperfect heuristics~\cite{megatronbridge} or trial-and-error to select an effective parallelization plan.

\subsection{Automatic LLM parallelizers}
\label{sec:background:auto_llm_parallelizers}

Automatic LLM parallelizers must search for a parallelization plan that results in high training throughput by exploring combinations of parallelisms and memory-saving techniques. As inputs, they take an LLM definition with the model architecture, and a GPU cluster configuration describing available resources and hardware specifications. They output a parallelization plan that specifies which parallelization strategies are applied and how, \eg MDP configurations, and how data/model partitions are placed across GPUs. The plan can be orchestrated for execution by a distributed LLM runtime, such as DeepSpeed~\cite{rasley2020deepspeed} or Megatron~\cite{shoeybi2019megatron}.

\T\ref{tab:autopara_survey} compares existing LLM parallelizers for distributed LLM training across three design dimensions:

\myparr{(i)~Supported techniques} determine the types of parallelization plans that can be generated. The more parallelisms and memory-saving techniques a parallelizer supports, the higher the potential training throughput of the created plans. Supporting more techniques, however, also increases the size of the search space (as shown in \F\ref{fig:search_space_incremental}). To balance search cost, different LLM parallelizers target specific use cases based on their supported techniques. For example, a parallelizer with CP support~\cite{miao2022galvatron, li2025hypertron, hetauto} is best-suited for long-sequence training, because CP offers little benefit for short sequences.


\myparr{(ii)~Search algorithms} determine how the LLM parallelizer explores combinations of parallelization techniques. When faced with large search spaces, parallelizers must use effective search algorithms to find high-throughput parallelization plans within a reasonable time budget. Existing LLM parallelizers (i)~explore intra- and inter-operator parallelisms hierarchically~\cite{zheng2022alpa, miao2022galvatron, zhu2025mist, um2024metis, harp}; and/or (ii)~rely on search heuristics, such as beam~\cite{liu2024aceso} or Monte Carlo tree search~(MCTS)~\cite{hetauto}.

\myparr{(iii)~GPU placement} strategies determine how to place data/model partitions from a parallelization plan across GPUs. The GPU placement specifies which GPUs perform which computation and how they communicate. For example, under DP, GPUs must synchronize gradients using all-reduce communication after processing assigned data partitions. Since GPUs may have different compute capabilities (FLOPs) and interconnect bandwidths, the placement strategy impacts training throughput for a given parallelization plan.

\subsection{Homogeneous vs.\ heterogeneous LLM parallelizers}
\label{sec:background:homo_het_parallelizers}

In \T\ref{tab:autopara_survey}, we categorize LLM parallelizers as \emph{homogeneous} LLM parallelizers, which assume homogeneous GPUs; and \emph{heterogeneous} LLM parallelizers, which account for GPU heterogeneity by performing GPU placement jointly with parallelization planning.

\mypar{Parallelisms} Homogeneous and heterogeneous LLM parallelizers both support the common parallelisms (DP, PP, and TP). More recently proposed parallelisms (CP and EP) are also supported by homogeneous parallelizers~\cite{zheng2022alpa, miao2022galvatron, li2025hypertron}, but ignored by heterogeneous LLM parallelizers or only supported partially~\cite{hetauto, harp}. Since the search space size with homogeneous GPUs is smaller, new parallelisms can be integrated without an explosion of search time, making heterogeneous LLM parallelizers more limited.

\mypar{Memory-saving techniques} The difference between homogeneous and heterogeneous LLM parallelizers becomes even more apparent for memory-saving techniques. Homogeneous LLM parallelizers~\cite{miao2022galvatron, liu2024aceso, zhu2025mist} at least partially consider memory-saving techniques in the search; in contrast, heterogeneous LLM parallelizers do not consider any of them, because, similar to the parallelisms, the explosion of the search space hinders the integration of memory-saving techniques.

\mypar{GPU placement} Homogeneous LLM parallelizers can implement a simple GPU placement heuristic: they select the correct number of GPU nodes indifferently, and only use DP and TP across GPU nodes when intra-operator parallelisms scale beyond the GPUs of a single node~\cite{zheng2022alpa, miao2022galvatron, liu2024aceso, zhu2025mist}. Instead, heterogeneous LLM parallelizers must search jointly for a good GPU placement and parallelization plan.

In heterogeneous GPU clusters, the different types of GPUs can cause compute imbalance and communication bottlenecks. Therefore, heterogeneous LLM parallelizers exploit uneven partitioning during GPU placement: parallelizers with \emph{uneven data partitioning}~\cite{hexiscale, um2024metis} assign different numbers of training batches to heterogeneous GPUs across DP model replicas---with more compute-capable GPUs being assigned more data to process. This suffers from increased memory usage, as it grows the batch size; parallelizers with \emph{uneven pipeline partitioning}~\cite{li2022amp, strati2025sailor, hexiscale, um2024metis, harp, hetauto} place pipeline stages of different sizes, \eg different numbers of layers, on heterogeneous GPUs. 

We conclude that existing heterogeneous parallelizers~\cite{li2022amp, strati2025sailor, hexiscale, um2024metis, harp, hetauto} fall short of supporting efficient distributed LLM training in heterogeneous GPU clusters. Although they address workload imbalances caused by GPU placements across heterogeneous GPUs, they do so by omitting parallelization techniques to maintain scalability. This prevents them from considering techniques that improve training throughput across all training scenarios, especially those that rely on memory-saving techniques (see \S\ref{sec:eval:het-skew}).


\section{\sys Design}
\label{sec:overview}

\begin{figure}[t]
  \centering
  \includegraphics[width=\columnwidth]{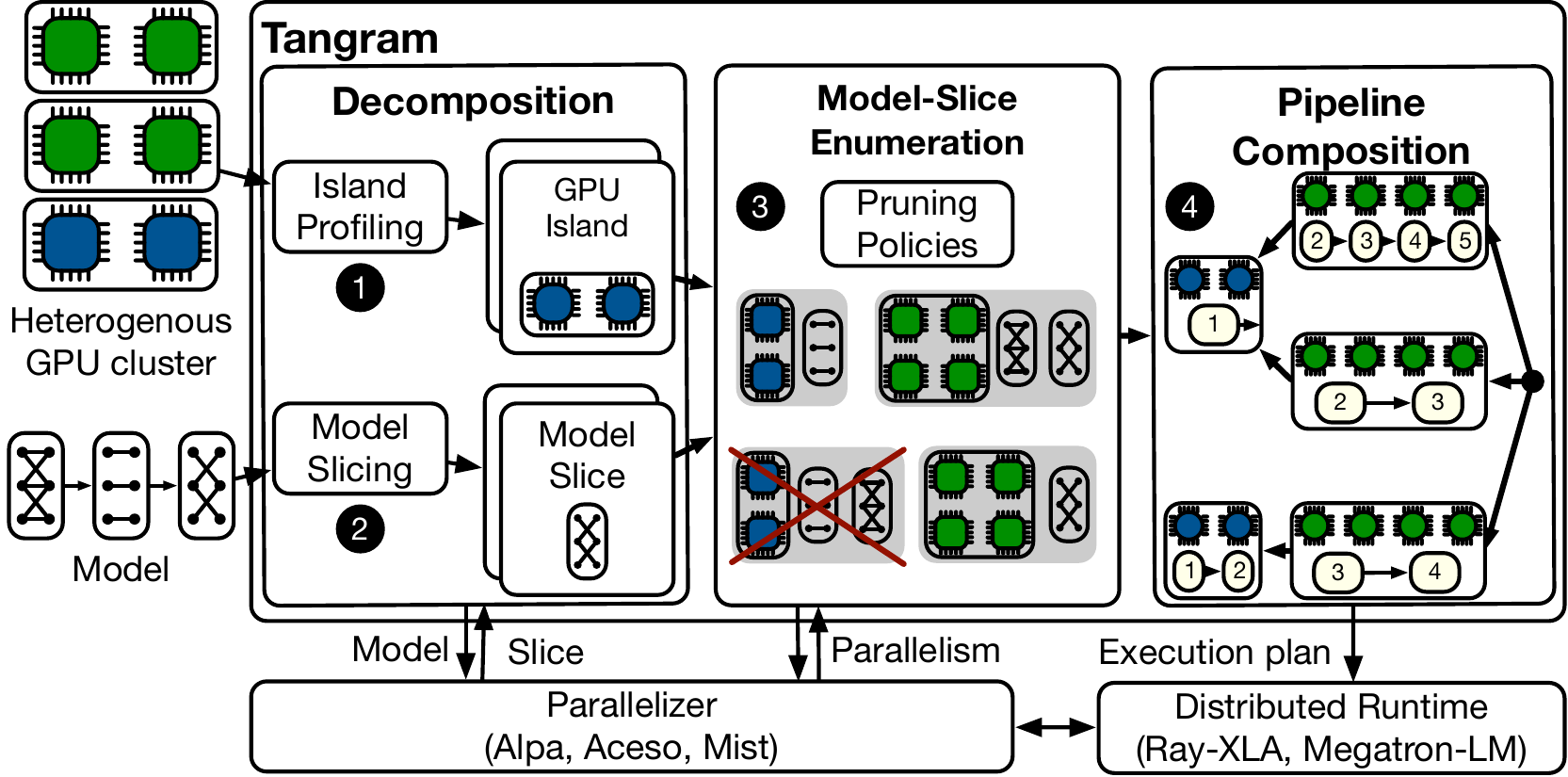}
  \caption{\sys design overview}
  \label{fig:overview}
\end{figure}

\sys decouples parallelization planning from GPU heterogeneity to take advantage of the full feature set of homogeneous LLM parallelizers in heterogeneous GPU clusters. As shown in \F\ref{fig:overview}, \sys{}'s design exploits two key ideas: (1)~heterogeneous GPU clusters have sets of GPUs with similar compute, memory, and connectivity resources; and (2)~existing homogeneous LLM parallelizers can commonly partition parts of a model.

With the help of a homogeneous LLM parallelizer, \sys produces a heterogeneous parallelization plan as follows:

\mypar{(1)~GPU resource and model decomposition~(\S\ref{sec:decomposition})} First, \sys constructs \emph{GPU islands}~(\myc{1}) by grouping GPU nodes whose GPU compute, memory, and intra-node connectivity are similar (\ie within a tunable threshold) and whose inter-node connectivity is high~(\S\ref{sec:cluster_decomposition}). To keep the number of islands low, \sys applies agglomerative clustering based on inter-node connectivity, because inter-node bandwidths can be asymmetric. After that, \sys decomposes the model into \emph{model slices}~(\myc{2})~(\S\ref{sec:model_slicing}), which form contiguous work units that \sys can assign to GPU islands. It first requests the smallest model partitions (\eg an attention layer in a dense LLM) from the LLM parallelizer through a generic \emph{parallelizer interface}. \sys then merges these model partitions into model slices of different sizes (\eg different numbers of layers) to pair with GPU islands. Each model-slice/island pair can be used as input to the LLM parallelizer.

\mypar{(2)~Model-slice enumeration~(\S\ref{sec:slice_plan_generation})}
Next, \sys enumerates \emph{model-slice/island pairs}, and queries the LLM parallelizer using them as inputs~(\myc{3}). An LLM parallelizer produces parallelization plans for each pair. To reduce exploration time, \sys prunes model-slice/island pairs using a set of policies: (i)~\emph{redundancy pruning} excludes \msi pairs with equivalent GPU islands and model slices; (ii)~\emph{imbalance pruning} excludes pairs that assign model slices with significant compute workload to resource-limited GPU islands, and vice versa; and (iii)~\emph{infeasibility pruning} excludes pairs that have no feasible parallelization plans due to memory constraints. In addition, \sys parallelizes the enumeration: each worker node executes LLM parallelizer queries over different subsets of model-slice/island pairs, reducing enumeration time.

\mypar{(3)~Pipeline composition~(\S\ref{sec:search_scalability})} \sys composes the model-slice/island plans~(\myc{4}) into a global heterogeneous plan. It constructs a pipeline in which each pipeline stage is a part of a model slice (\ie a sequence of layers) and is executed on GPUs from a single GPU island. As we show in~\S\ref{sec:search_scalability}, a model-slice/island plan can include multiple pipeline stages. The pipeline composition is performed by a dynamic programming algorithm, which creates sequences of model-slice/island pairs and associated parallelization plans, and finds the sequence with the lowest estimated iteration time. These sequences are memoized, \ie stored in a cache table, to avoid redundant traversals. Additionally, \sys identifies resource-limited GPU islands that would increase the iteration time and therefore introduce a hardware bottleneck. \sys excludes such GPU islands from the composed pipeline to avoid degrading the overall throughput.


\section{Resource and Model Decomposition}
\label{sec:decomposition}

Next, we describe how GPU islands are constructed~(\S\ref{sec:cluster_decomposition}) and how models can be decomposed into model slices using a generic LLM parallelizer interface~(\S\ref{sec:model_slicing}).

\subsection{GPU islands}
\label{sec:cluster_decomposition}

\mypar{Island definition} A GPU island must meet the following requirements: (R1)~it must constitute a set of resources that is a compatible input to an LLM parallelizer, \ie a group of GPU nodes with the same number of GPUs, and faster intra-node than inter-node bandwidth; (R2)~the resources of a GPU island must be homogeneous; and (R3)~communication between islands must not be a severe bottleneck.


We observe that resource heterogeneity in GPU clusters is not arbitrary: GPU clusters are often composed of sets of GPU nodes that were acquired at the same time in bulk. These nodes have GPUs of similar performance characteristics and dedicated interconnects~\cite{nvidiadgx}.

Let $\mathcal{G}$ be the set of GPUs in the cluster. Each GPU, $g \in \mathcal{G}$, belongs to a node~$n_g$ and is characterized by its compute capacity~$c_g$, memory capacity~$m_g$, and the bandwidth~$b_g$ of its intra-node interconnect. For GPUs $g$ and $g'$ on different nodes, $e(g,g')$ denotes the network bandwidth between $n_g$ and~$n_{g'}$. Given tolerances $\boldsymbol{\epsilon}=(\epsilon_c,\epsilon_m,\epsilon_i)$ for compute~$\epsilon_c$, memory~$\epsilon_m$, and intra-node interconnect~$\epsilon_i$, we define a GPU island as a subset~$I \subseteq \mathcal{G}$ of unique GPUs that satisfies the following constraints:

\begin{myenumerate}
    \item \textbf{Compute/memory homogeneity.} All nodes in $I$ must have the same number of GPUs to satisfy compatibility~(R1), and for each $g, g' \in I$, compute and memory capacities must be sufficiently homogeneous to avoid compute underutilization~(R2):
    \begin{equation}
        \frac{\max(c_g, c_{g'})}{\min(c_g, c_{g'})} \leq 1+\epsilon_c \quad\quad   \frac{\max(m_g, m_{g'})}{\min(m_g, m_{g'})} \leq 1+\epsilon_m
    \end{equation}
    \item \textbf{Intra-node network parity.} For each $g, g' \in I$, the intra-node interconnects of the host nodes must have similar bandwidth to bound communication bottlenecks within the island~(R2):
    \begin{equation}
        \frac{\max(b_g, b_{g'})}{\min(b_g, b_{g'})} \leq 1+\epsilon_i
    \end{equation}
    \item \textbf{Inter-node network proximity.} For each $g, g' \in I$ in different nodes and each $g'' \notin I$, links within the GPU island must be at least as fast as links leaving it:
    \begin{equation}
        e(g,g') \geq e(g,g'')
    \end{equation}
    This condition allows some parallelization strategies that require heavy communication, such as EP, to apply across nodes of a GPU island~(R3). Across islands, \sys must only choose plans with limited peer-to-peer data transfers (\eg across PP stages) to avoid bottlenecks.
\end{myenumerate}

\begin{table*}[t]
  \centering
  \small
  \begin{tabular}{@{}l p{0.46\linewidth}@{}}
    \toprule
    \textbf{Function} & \textbf{Behavior} \\
    \midrule
    \texttt{slice\_model(model) -> list(model\_slice)}
                      & Partitions model into sequence of model atoms \\[2pt]
    \texttt{concatenate(slice\_1, slice\_2) -> model\_slice}
                      & Concatenate two model slices \\[2pt]
    \texttt{profile(slice, island) -> performance\_profile}
                      & Retrieve profiling results for a model slice on a GPU island \\[2pt]
    \texttt{parallelize(slice, island, $\ldots$) -> model\_slice\_plan}
                      & Parallelize a model slice on a GPU island \\[2pt]
    \bottomrule
  \end{tabular}
  \caption{\sys parallelizer interface \textnormal{(non-exhaustive)}}
  \label{tab:interface}
\end{table*}

\mypar{Island construction} After profiling each GPU node for the relevant performance metrics, \sys: \myc{1} decomposes the cluster into groups of GPU nodes whose compute and memory spreads are within $\epsilon_c$ and $\epsilon_m$; \myc{2} decomposes each group by separating nodes with intra-node bandwidth spreads larger than $\epsilon_i$; and, in the remaining groups, \myc{3} identifies GPU nodes that have high inter-node bandwidth between them. To keep the number of islands low in step \myc{3}, \sys greedily merges nodes using agglomerative clustering~\cite{agglomerativeclustering}. The output consists of disjoint groups of GPU nodes that satisfy the conditions for GPU islands.

In typical heterogeneous GPU clusters, the number of GPU islands is often small, \eg 2--6. Organizations operate only a few generations of GPUs at a time and retire old GPUs regularly~\cite{MLaaS, bootseer}. Across generations, both compute and interconnects differ significantly~\cite{dgx2017nvidia, h100, blackwell, nvidiadgx}. This makes \sys's GPU island construction robust in practice and does not result in excessive fragmentation, even with limited cluster heterogeneity~\cite{sai2024kalos}.

\subsection{Model decomposition}\label{sec:model_slicing}

\sys uses the LLM parallelizer to parallelize parts of the model on GPU islands. Since an ineffective distribution of the workload can lead to resource contention on GPU islands, \sys must balance the amount of work assigned to each island to ensure high resource utilization. When assigning workload to islands, \sys slices the model into model slices and pairs each slice with a GPU island.

\mypar{Model slicing} A model slice is a contiguous model partition. A challenge is that each LLM parallelizer uses its own model representation and thus determines the granularity and level (\eg full graph or forward-only layers) of a model partition. For example, Alpa~\cite{zheng2022alpa} defines a model as a JAX~\cite{jax2018github} computational graph and partitions the model at the granularity of layers, which consist of many operators; Aceso~\cite{liu2024aceso} partitions the model at the granularity of operators. A model slice thus must encapsulate approaches across parallelizers.

\sys exploits the insight that LLM parallelizers have similar APIs that support model partitioning and concatenation. Such APIs can be queried by \sys to expose how a model can be partitioned. Using the model definition as input to the model partitioning API, \sys obtains the smallest partitions of a model, and encapsulates these as \emph{model atoms}. They represent the most granular units of the computational workload. A concatenation of model atoms forms a model slice. \sys does not need knowledge of the internals of a model atom for planning. By assigning model slices constructed from model atoms, \sys balances workload across GPU islands.

Model atoms can be coarse-grained to reduce the planning time, but this may lead to imbalanced workloads across GPU islands. If model atoms are as fine-grained as the smallest partition provided by the LLM parallelizer, \sys can balance the workload better across islands at the cost of more planning time. We choose the granularity empirically---in \S\ref{sec:evaluation}, we show that using model layers as model atoms results in effective plans within reasonable search time.

\mypar{Parallelizer interface} \sys defines a generic interface for LLM parallelizer integration (see~\T\ref{tab:interface}). The interface exposes four high-level functions that parallelizers implement internally. Since parallelizers commonly partition a model and combine model partitions to explore different partitioning schemes, \sys uses these functions directly as (1)~\code{slice\_model()} to partition an LLM into model atoms; and (2)~\code{concatenate()} to combine model slices. Since parallelizers must gather profiling information before parallelizing a model on a GPU cluster, \sys also invokes the \code{profile()} and \code{parallelize()} functions to obtain profiling results for GPU islands and parallelization plans, providing a model-slice/island pair as input.
    
Due to the interface's simplicity, a new LLM parallelizer can be added with low effort. For example, the parallelizer interface can be implemented for Alpa by reusing functions in its inter-operator compilation passes to obtain a model-slice/island plan~\cite{zheng2022alpa}.

\section{Model-Slice Composition}
\label{sec:recomposition}

In this section, we describe how \sys enumerates model-slice/island pairs with pruning~(\S\ref{sec:slice_plan_generation}), and how it creates a heterogeneous plan by composing partial plans~(\S\ref{sec:search_scalability}).

\subsection{Pruning model-slice enumeration}
\label{sec:slice_plan_generation}

\begin{figure}[t]
    \centering
    \includegraphics[width=\columnwidth]{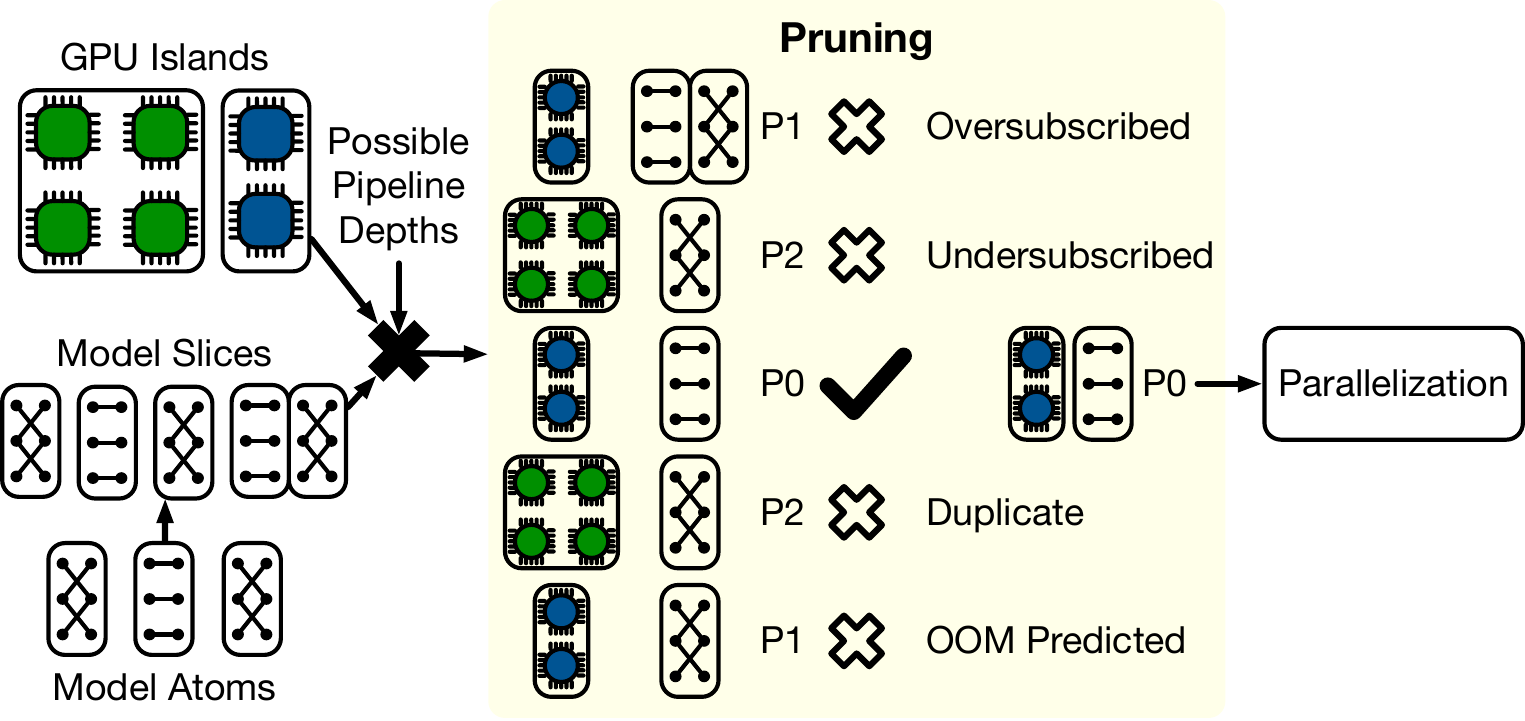}
    \caption{Pruning \msi enumeration}
    \label{fig:pruning}
\end{figure}

For a heterogeneous plan with high training throughput, \sys must balance the workload across GPU islands while avoiding underutilization. \sys enumerates potential \msi pairs by parallelizing the model slice for each island using an LLM parallelizer. The number of pairs grows exponentially with the number of GPU islands and model size, making naive enumeration infeasible. Therefore, \sys prunes model-slice/island pairs, queries the parallelization plans in parallel, and discards infeasible plans early (see \F\ref{fig:pruning}).

Given a model architecture and heterogeneous GPU cluster, many model-slice/island pairs are redundant, infeasible, or underutilize resources. \sys prunes such pairs using three policies:

\mypar{(i)~Redundancy pruning} \sys exploits the repetitive structure of models (\eg repeated transformer or MoE layers) to avoid enumerating duplicate model slices on the same GPU island (see~\F\ref{fig:pruning}; duplicates are discarded). During model slicing, the LLM parallelizer provides a signature of model atoms (\eg a hash value). \sys uses the signature to identify model slices with the same composition, and invokes the LLM parallelizer only once.

\mypar{(ii)~Imbalance pruning} \sys prunes \msi pairs that would imbalance the workload. A model slice causes a GPU island to stall if it requires more compute than the island's capacity relative to the cluster's compute capacity; the opposite leads to underutilization (see~\F\ref{fig:pruning}; \msi pairs that cause over- and under-subscription are discarded). For each GPU island (see~\S\ref{sec:cluster_decomposition}), \sys estimates its relative compute capacity by profiling throughput of model atoms. \sys can then estimate compute demand for model slices based on their number of atoms. To avoid underutilization, \sys prunes the \msi pairs when compute capacity and demand are mismatched. Since throughput depends on intra-operator parallelization (\eg CP and TP), \sys uses profiling to estimate the compute capacity of each GPU island.

\mypar{(iii)~Infeasibility pruning} \sys removes pairs that cannot be parallelized due to memory constraints. If a model slice is too large to fit in the GPU island's aggregate memory, \sys discards the corresponding pair (see \F\ref{fig:pruning}; \msi pairs that overflow memory are discarded). \sys estimates the minimum memory requirement of a model slice based on the parallelizer's profiling information. 

\tinyskip

\noindent
Due to requirement~(R1) of GPU islands~(see \S\ref{sec:cluster_decomposition}), each LLM parallelizer query on a distinct model-slice/island pair is independent. This makes model slice parallelization itself parallelizable: \sys distributes queries to different workers, further reducing generation time.


\subsection{Pipeline composition}
\label{sec:search_scalability}

Enumerating model-slice/island pairs results in a comprehensive set of partial plans with estimated execution times and memory requirements. \sys composes these into a \emph{global} heterogeneous plan using pipeline parallelism~(PP)~\cite{huang2019gpipe,narayanan2019pipedream}. PP has three advantages: (1)~it is communication-efficient, only transferring stage inputs and outputs (\ie activations) between stages on potentially bandwidth-limited interconnects; (2)~it is memory-efficient by not replicating data or model parameters; (3)~it is orthogonal to other parallelisms, and pipeline stages can be parallelized independently.

\begin{figure}[t]
  \centering
  \includegraphics[width=\columnwidth]{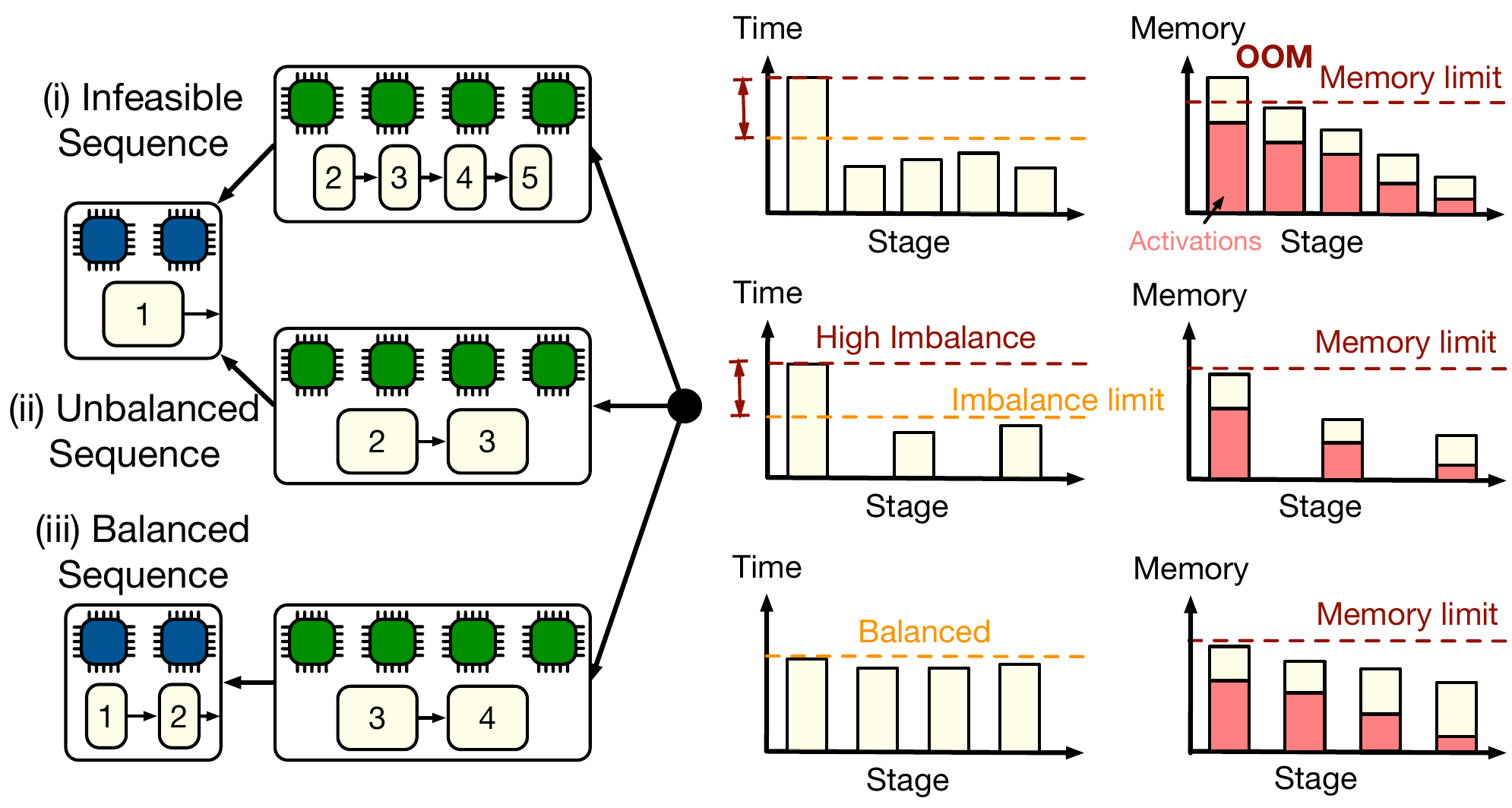}
  \caption{Pipeline plan search \textnormal{(Example \msi sequences that show differences in feasibility and performance.)}}
  \label{fig:search-tree}
\end{figure}

Under PP, the memory usage of a pipeline stage depends on its position within the pipeline~\cite{fan2021dapple}~(see \F\ref{fig:search-tree}). The deeper the pipeline, the more activations (\ie stage inputs and outputs) are held by earlier stages in memory before being released in the backward pass. Given a model-slice/island plan, a GPU island can only accommodate a fixed number of subsequent stages; otherwise, it would require a new parallelization plan with extra memory usage. Instead of requesting model-slice/island plans for each pipeline position and depths, \sys reuses a plan until it is no longer feasible. Then, \sys requests a new parallelization plan from the LLM parallelizer, as long as GPU memory has not yet been exhausted. This approach reduces the number of plans needed for each model-slice/island pair, and thus enumeration time.

\sys composes partial plans into a pipeline with \emph{balanced} stages. It uses a 1F1B pipeline~\cite{narayanan2019pipedream}, whose schedule bounds in-flight activations by the pipeline depth, which reduces peak memory usage. A challenge with composition is that the number of model-slice/island plans is potentially large and that model-slice/island plans themselves can contain several pipeline stages depending on the output of the LLM parallelizer. \sys therefore uses a dynamic programming approach, which iteratively constructs sequences of model-slice/island pairs to compose a workload-balanced pipeline for the global plan.

\begin{figure*}[t]
  \centering
  \includegraphics[width=\textwidth]{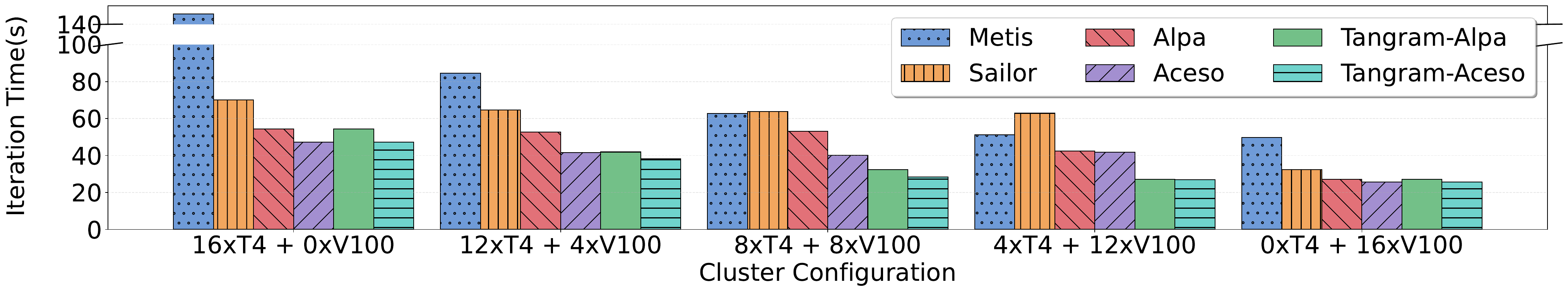}
  \caption{Training iteration time with varying heterogeneous GPU ratios \textnormal{(GPT-3 $6.7$B model with two GPU types in a 16-GPU cluster, MBS=1)}}
  \label{fig:hetml_skew}
\end{figure*}

The algorithm works as follows: (1)~it ranks the slowest stages in all model-slice/island plans and picks the one with the lowest execution time. because pipeline throughput is dominated by the slowest stage. That stage represents the maximum stage imbalance permitted, \ie bottleneck, at the current composition step; (2)~it greedily constructs sequences of model-slice/island plans to cover the full model without adding a stage that introduces a new bottleneck, which would make the sequence more unbalanced. Model-slice/island sequences are constructed by adding new \msi in front of the sequence, which enables memoization and accounts with the increase in memory usage due to activations; (3)~it compares the best global plan found so far with the best sequence built in the current composition step, and potentially updating the global plan; and (4)~it repeats previous steps while gradually relaxing the constraints, \ie allowing a larger stage imbalance in the composition, until no further pipeline latency improvement is possible.

\F\ref{fig:search-tree} illustrates the memoization of \msi sequences and their construction under the stage imbalance constraint. As model-slice/island plans are added, a sequence can become (i)~infeasible, because the current number of subsequent stages exhausts memory, or become (ii)~unbalanced, because the plan contains a pipeline stage with a new dominant bottleneck. \sys discards such sequences and memoizes the best (iii)~feasible ones.

Note that \sys can eliminate a resource-limited GPU island and use a subset of the GPU cluster. If a resource-limited GPU island would always introduce a severe bottleneck with any model-slice/island plan, it is eliminated from the heterogeneous plan during pipeline composition.



\section{Evaluation}
\label{sec:evaluation}

We evaluate \sys in various scenarios with heterogeneous GPU nodes that have compute, memory, and network heterogeneity. Our goal is to answer the these questions:

\begin{myitemize}
  \item Does \sys compose high-throughput heterogeneous plans for training different LLMs on heterogeneous GPU clusters? (\S\ref{sec:eval:het-skew})
  \item How effective is \sys in using existing homogeneous LLM parallelizers? (\S\ref{sec:eval:homogeneous})
  \item Can \sys scale LLM training with heterogeneous GPUs? (\S\ref{sec:eval:heterogeneity})
  \item Do \sys's pruning policies speed up the planning process? (\S\ref{sec:eval:pruning}) 
  \item How scalable is \sys as GPU clusters become more heterogeneous? (\S\ref{sec:eval:search})
\end{myitemize}

\subsection{Experimental setup}
\label{sec:eval:setup}

Our experiments have the following setup:

\mypar{GPU clusters} We use 5~types of Nvidia GPU instances from Azure Cloud virtual machines~(VMs) to construct three heterogeneous training clusters within the same region: \textsf{C1-compute-het}, \textsf{C2-modern-het} and \textsf{C3-full-het} (see~\T\ref{tab:exp_env}), which differ in compute performance, memory capacity, and intra-node interconnect bandwidth.

\mypar{Training workloads} We evaluate \sys on both dense and MoE LLMs: for dense LLMs, we use GPT-3~\cite{brown2020language} with a sequence length of 2048; for MoEs, we use GShard MoEs~\cite{lepikhin2020gshard} with a sequence length of 1024 for comparing against Alpa, and Qwen3-MoE~\cite{qwen3} and Mixtral MoEs~\cite{mixtral} with a sequence length of 4096 for integrating with new parallelization strategies from Galvatron~\cite{miao2022galvatron}. For all LLMs, we use a global batch size of 128 and mixed-precision training with FP16 activations to reduce the memory footprint.

\mypar{Baselines} To compare with state-of-the-art LLM parallelizers, we use Aceso~\cite{liu2024aceso} and Alpa~\cite{zheng2022alpa} as homogeneous LLM parallelizers, Metis~\cite{um2024metis} and Sailor~\cite{strati2025sailor} as heterogeneous LLM parallelizers. For training with Qwen3 and Mixtral MoEs, we implement a simple homogeneous LLM parallelizer on top of Galvatron~\cite{miao2022galvatron}, which enumerates parallelization strategies (including EP) with uniform partitioning. We make minor changes to Alpa's and Sailor's memory estimation logic to reduce overestimation in heterogeneous GPU settings. For Aceso, we set its search budget to $100$~seconds on heterogeneous GPU clusters to balance search time against plan quality. For Metis, we integrate its planner with Aceso's runtime and memory estimation models for a fair comparison with \sys. We use Sailor's throughput-focused mode and deactivate its plan filtering based on GPU node costs.

\begin{table}[t]
  \centering
  \footnotesize
  \setlength{\tabcolsep}{3pt}
  \caption{Evaluation GPU clusters \textnormal{(\textsf{C1-compute-het}: same memory+mixed compute/interconnect; \textsf{C2-modern-het}: modern high-end GPUs; \textsf{C3-full-het}: full heterogeneity)}}
  \begin{tabular}{l c c c c c}
    \toprule
    \multirow{3.4}{*}{\textbf{GPU type}} &
    \multirow{3.4}{*}{\textbf{Mem.}} &
    \multirow{3.4}{*}{\shortstack[c]{\textbf{Intra-node}\\\textbf{link}}} &
    \multicolumn{3}{c}{\textbf{\# GPUs per cluster}} \\
    \cmidrule(lr){4-6}
    & & & \shortstack[c]{\textbf{C1}\\\scriptsize compute-het}
          & \shortstack[c]{\textbf{C2}\\\scriptsize modern-het}
          & \shortstack[c]{\textbf{C3}\\\scriptsize full-het} \\
    \midrule
    T4         & 16\,GB & PCIe   & 4$\times$4 & --  & 1$\times$4 \\
    V100-16    & 16\,GB & PCIe   & 4$\times$4 & --  & 1$\times$4 \\ 
    V100-32    & 32\,GB & NVLink & -- & --  & 1$\times$8  \\
    A100-80    & 80\,GB & NVLink & -- & 2$\times$4   & 1$\times$4  \\
    H100\,NVL  & 94\,GB & NVLink & -- & 4$\times$2   & -- \\ 
    \midrule
    \textbf{Total} & & &  \textbf{32} & \textbf{16} & \textbf{16} \\
    \bottomrule
  \end{tabular}
  \label{tab:exp_env}
\end{table}

\subsection{Parallelization with heterogeneous GPUs}
\label{sec:eval:het-skew}

\mypar{Dense LLMs} We investigate how \sys performs for training dense LLMs with different GPU types. In this experiment, we compare the parallelization plans for different GPUs inside the heterogeneous cluster~\textsf{C1-compute-het}. We vary the ratio of T4 to V100 GPUs between 12:4 to 8:8 and 4:12. In each case, \sys uses two GPU islands with the respective GPU types. We measure the training iteration time of GPT-3~6.7\unit{B} for each system with 16~GPUs.

As \F\ref{fig:hetml_skew} shows, \sys consistently outperforms the baselines. The speedups over Metis, Sailor, Alpa, and Aceso are, on average, $2.1\times$, $2.1\times$, $1.6\times$, and $1.4\times$, respectively. This improvement stems from \sys{}'s ability to leverage the full set of parallelization strategies from the underlying LLM parallelizers and composing a pipeline of balanced stages. The larger improvements over Metis and Sailor are because they do not support activation recomputation, but it enables V100 GPUs to execute a larger model partition. Instead, Metis and Sailor must use micro-batch size~(MBS) of 1, and rely on TP to reduce the per-GPU memory usage. Metis also does not compute the same plan as Sailor, because its memory estimation declares plans infeasible. Across all setups, Sailor generates the same uniform PP partitioning plan, independent of the number of V100 GPUs. Sailor's iteration time plateaus, because the pipeline latency becomes bottlenecked by stages assigned to T4 GPUs.

Based on \sys's training throughput, we conclude that decoupling parallelization planning from heterogeneity and our pruning policies (\S\ref{sec:slice_plan_generation}) do not negatively affect the quality of the solutions found. In addition, \sys can use the most-suitable parallelizer in this scenario (Aceso) to achieve the highest training throughput.

\begin{figure}[t]
  \centering
  \includegraphics[width=\columnwidth]{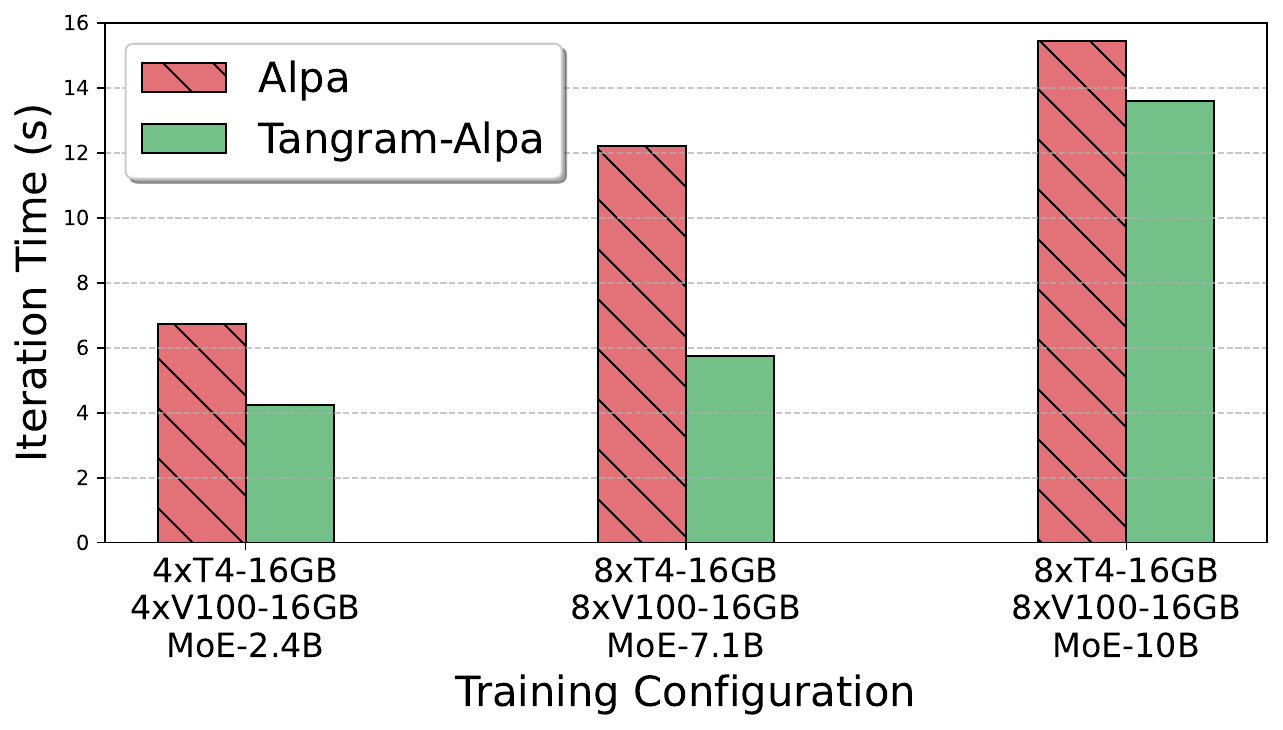}
  \caption{Training iteration time for MoE models \textnormal{(GShard MoE model with varying sizes)}}
  \label{fig:hetml_moe}
\end{figure}

\mypar{MoE Models}
We validate that \sys generates high-throughput heterogeneous plans for multiple LLM architectures, benefiting from the latest GPU hardware and parallelization strategies. We evaluate \sys with GShard MoE models, which have been supported by Alpa~\cite{zheng2022alpa}, and newer Mixtral~\cite{mixtral} and Qwen3-MoE models~\cite{qwen3}, which current LLM parallelizers do not support.

First, we compare \sys-Alpa against Alpa on GShard MoE models across three sizes ($2.4$B, $7.1$B, and $10$B) on GPUs from cluster~\textsf{C1-compute-het}. \F\ref{fig:hetml_moe} reports the iteration time of \sys-Alpa and Alpa. We use ratios between T4 and V100 GPUs of 4:4 and 8:8. For a size of $2.4$B, Alpa has an iteration time of $6.7$\unit{s} compared to \sys-Alpa{}'s $4.2$\unit{s}, showing a speedup of $1.6\times$; for a size of $7.1$B, the times are $12.2$\unit{s} and $5.8$\unit{s} respectively, which is a $2.1\times$ speedup. For the largest model ($10$B), Alpa's time is $15.5$\unit{s} and \sys-Alpa's time is $13.6$\unit{s} ($1.1\times$ speedup). Alpa's approach discovers plans similar to EP. In contrast, Aceso would require a manual implementation of parallel operators for new LLMs, which is not available for MoE models. Neither Sailor nor Metis incorporate the ability to search over the EP dimension. \sys exploits GShard EP support through Alpa, thus demonstrating the benefit of \sys's decoupled design.

With $7.1$B parameters, where the greatest improvement is observed, Alpa utilizes the whole GPU cluster. \sys-Alpa identifies the opportunity to eliminate the performance bottlenecks of the T4 GPUs by using a single-island V100-only plan, which results in lower iteration time. With the larger $10$B model, the limited memory capacities of T4 and V100 GPUs become the new bottleneck, which prevents \sys from balancing the computational workload across heterogeneous GPUs. Since \sys can interface with different LLM parallelizers, a homogeneous LLM parallelizer that explores EP jointly with additional memory-saving techniques could yield further improvements.

\begin{figure}[t]
  \centering
  \includegraphics[width=\columnwidth]{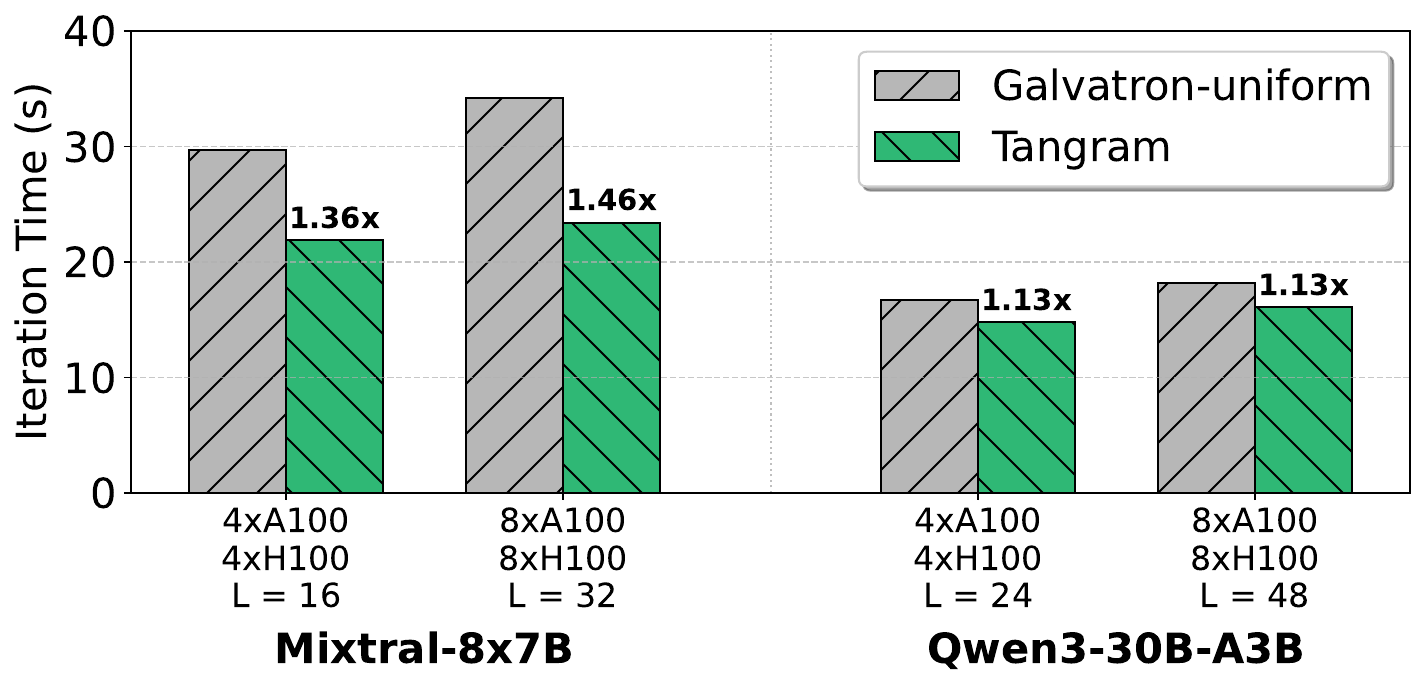}
  \caption{Training iteration time for recent MoE models with Galvatron's FSEP \textnormal{(Mixtral-8$\times$7B and Qwen3-30B-A3B, trimmed and full layer counts on cluster \textsf{C2-modern-het}, MBS=4)}}
  \label{fig:tangram_laer}
\end{figure}

To train Qwen3-30B-A3B and Mixtral-8$\times$7B, and their trimmed versions with half the number of layers, we use cluster~\textsf{C2-modern-het} with Hopper-class GPUs~\cite{h100}. We executes plans in the Galvatron MoE runtime~\cite{miao2022galvatron}, which uses FlashAttention-2~\cite{flashattention2} and combines EP and FSDP~\cite{pytorchfsdp}, called \emph{fully-sharded expert parallelism}~(FSEP)~\cite{laermoe}. FSEP fully partitions the parameters of each expert by the number of expert-parallel GPUs for better token routing, expert placement, and memory savings~\cite{pytorchfsdp}. Since the Galvatron's LLM parallelizer currently does not consider EP in its search space, we implement a naive homogeneous LLM parallelizer, called \textsf{Galvatron-uniform}, that enumerates plans with uniform DP, TP, PP and FSEP partitioning.

\F\ref{fig:tangram_laer} shows the iteration time of the plans produced by \sys and the LLM parallelizer. For Mixtral-8$\times$7B, \sys reduces the iteration time from $30$\unit{s} to $22$\unit{s} ($1.4\times$) for the trimmed 16-layer model, and from $34$\unit{s} to $23$\unit{s} ($1.5\times$) for the full 32-layer model. For Qwen3-30B-A3B, \sys improves performance marginally from $17$\unit{s} to $15$\unit{s} and from $18$\unit{s} to $16$\unit{s}, a $1.1\times$ speedup in both cases. The improvement is due to \sys's ability to compose a heterogeneous plan that constructs a high-throughput pipeline with balanced stages, assigning more layers to the H100 GPUs, instead of treating all GPUs the same. We also observe a significant slowdown in iteration time on the 32-layer Mixtral MoE when placing the A100 GPUs at the front of the pipeline, to $54$\unit{s}, caused by the increased memory pressure. Since Mixtral MoE has significantly more compute-intensive experts than the Qwen3 MoE, the gap in training throughput against the naive parallelizer becomes more pronounced.

These experiments show that \sys improves training throughput for different LLM architectures, while leveraging the latest LLM parallelizers, GPU hardware, and parallelization strategies. \sys also benefits from orthogonal techniques, such as FlashAttention-2, due to its ability to support different runtimes and combining them with LLM parallelizers.

\subsection{Parallelization with homogeneous GPUs}
\label{sec:eval:homogeneous}

To explore the generality of \sys{}'s approach, we also evaluate it on homogeneous GPU clusters. Without GPU heterogeneity, the drawbacks of omitting parallelization strategies is directly exposed and cannot be compensated by balancing training workload across heterogeneous GPUs. We use the same cluster~\textsf{C1-compute-het}, but only consider the T4 and V100 GPUs separately, yielding two homogeneous clusters.

With the T4 GPUs~(\F\ref{fig:hetml_skew}), Alpa and \sys-Alpa both have an iteration time of $54$\unit{s}; Aceso's and \sys-Aceso's iteration times are $47$\unit{s}. \sys reproduces the result of its underlying LLM parallelizers, because it only identifies a single GPU island. In constrst, Metis produces a plan with an iteration time of $146$\unit{s}, and Sailor has a time of $70$\unit{s}. \sys improves performance by $3.1\times$ over Metis and $1.5\times$ over Sailor, because it leverages all parallelization strategies supported by Alpa and Aceso, specifically activation recomputation~\cite{chen2016training}. Due to the more limited connectivity of T4 GPUs, heterogeneous LLM parallelizers favor plans with a low TP degree due to their lower communication overhead. The best parallelization plan produced by Aceso is a 16-stage PP plan, which is only viable with activation recomputation, explaining the worse performance of Metis and Sailor.

With V100 GPUs~(\F\ref{fig:hetml_skew}), \sys matches the homogeneous LLM parallelizers as well. \sys-Alpa has an iteration time of $27$\unit{s} and \sys-Aceso has $26$\unit{s}. For the heterogeneous LLM parallelizers, Metis has a $50$\unit{s} iteration time ($1.9\times$), and Sailor has $32$\unit{s} ($1.2\times$). Sailor's and Metis' slowdowns due to the use of TP become less significant compared to the T4 GPUs because of the faster interconnect between V100 GPUs. \sys produces parallelization plans that are similar to the ones generated by homogeneous LLM parallelizers, avoiding performance degradation.

\subsection{Scaling-up training with heterogeneous GPUs}
\label{sec:eval:heterogeneity}

\begin{figure}
  \centering
  \includegraphics[width=\columnwidth]{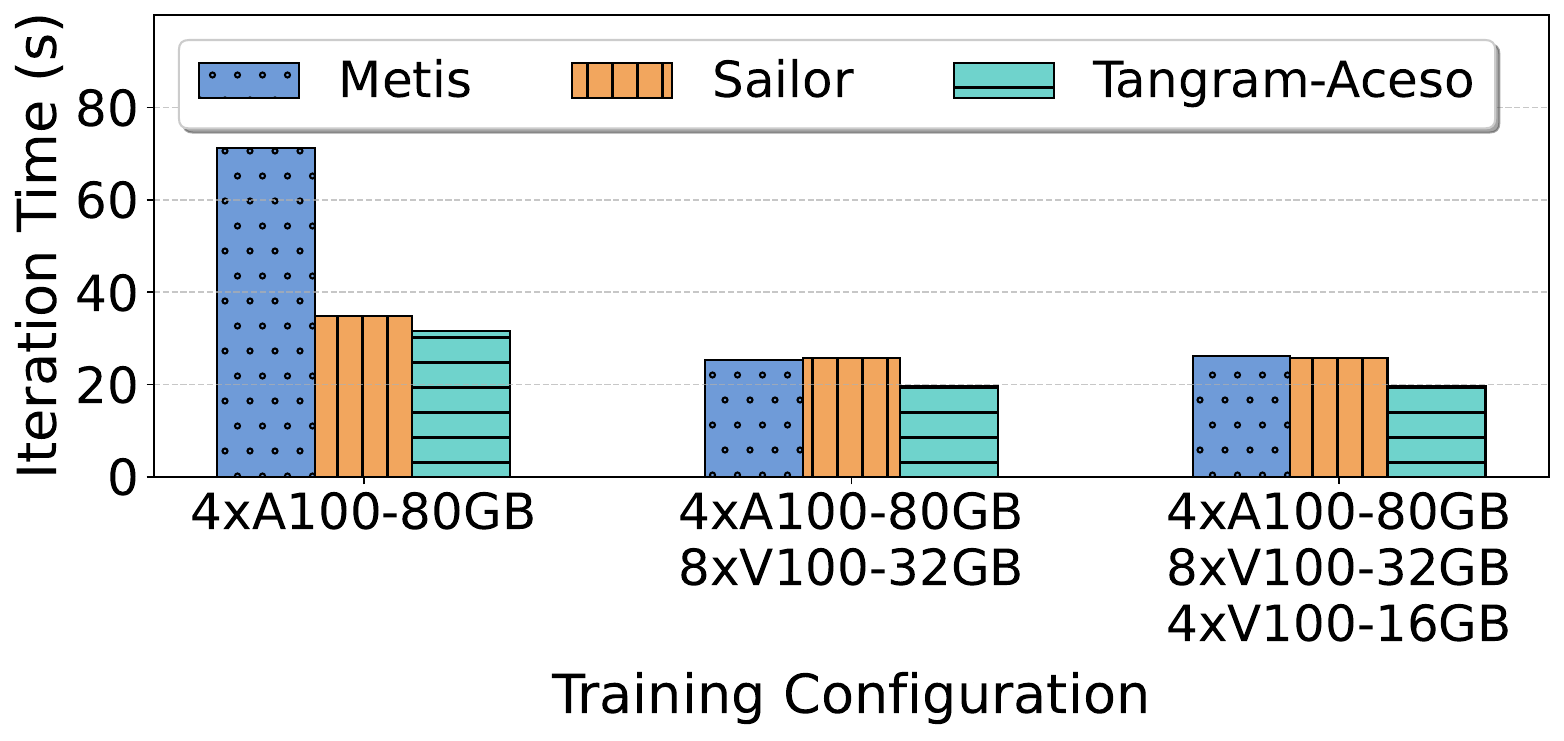}
  \caption{Training iteration time for different degrees of heterogeneity \textnormal{(GPT-3 $13$B model, MBS=1, cluster C3-full-het)}}
  \label{fig:hetml_island_scaling}
\end{figure}

In this experiment, we gradually add more heterogeneous GPUs to the cluster to train a 40-layer GPT-3 model with $13$B parameters. We measure the iteration time of Sailor, Metis, and \sys-Aceso, in the cluster~\textsf{C3-full-het}. The cluster configurations are (i)~4$\times$A100 GPUs; (ii)~4$\times$A100 GPUs with 8$\times$V100-32\unit{GB} GPUs; and (iii)~4$\times$A100 GPUs with 8$\times$V100-32\unit{GB} GPUs and 4$\times$V100-16\unit{GB} GPUs.

In the first configuration~(\F\ref{fig:hetml_island_scaling}), Metis has an iteration time of $71$\unit{s}, Sailor has $35$\unit{s}, and \sys-Aceso reduces this to $32$\unit{s}---a speedup of $2.2\times$ over Metis, and $1.1\times$ over Sailor. The second configuration adds a node with 8$\times$V100-32\unit{GB} GPUs. \F\ref{fig:hetml_island_scaling} shows that all systems achieve a noticeable speedup: $2.8\times$ for Metis, $1.3\times$ for Sailor, and $1.6\times$ for \sys-Aceso. Iteration times fall to $25$\unit{s}, $26$\unit{s}, and $20$\unit{s}, respectively. \sys's improvement over Metis is $1.3\times$ and $1.3\times$ for Sailor.

We analyze the generated heterogeneous plans: (i)~Sailor and Metis identify an MBS of 1 as the best, effectively disallowing heterogeneous DP; (ii)~neither considers inconsistent node sizes when composing the pipeline plan and would need to divide the 8-GPU node into two smaller nodes manually; and (iii)~both have a higher iteration time, because activation recomputation allows \sys-Aceso to reduce memory use. In addition, Metis has a slightly lower iteration time than Sailor due to its heterogeneity-aware pipeline compared to Sailor's uniform partitioning.

The third configuration adds a node with 4$\times$V100-16\unit{GB} GPUs. As \F\ref{fig:hetml_island_scaling} shows, the iteration times do not change, which implies that the 4$\times$V100-16\unit{GB} GPUs become a performance bottleneck. Both Sailor and \sys can explore partial cluster assignments to avoid this bottleneck. In contrast, Metis only considers parallelization strategies that utilize the entire cluster, resulting in degraded performance. Further investigation reveals that the bottleneck is due to slower inter-node communication. This showcases \sys's ability to identify bottlenecks during the pipeline plan composition~(\S\ref{sec:search_scalability}) and remove them from the global plan.

\subsection{Benefits of pruning policies}
\label{sec:eval:pruning}

\begin{figure}[t]
  \centering
  \includegraphics[width=\columnwidth]{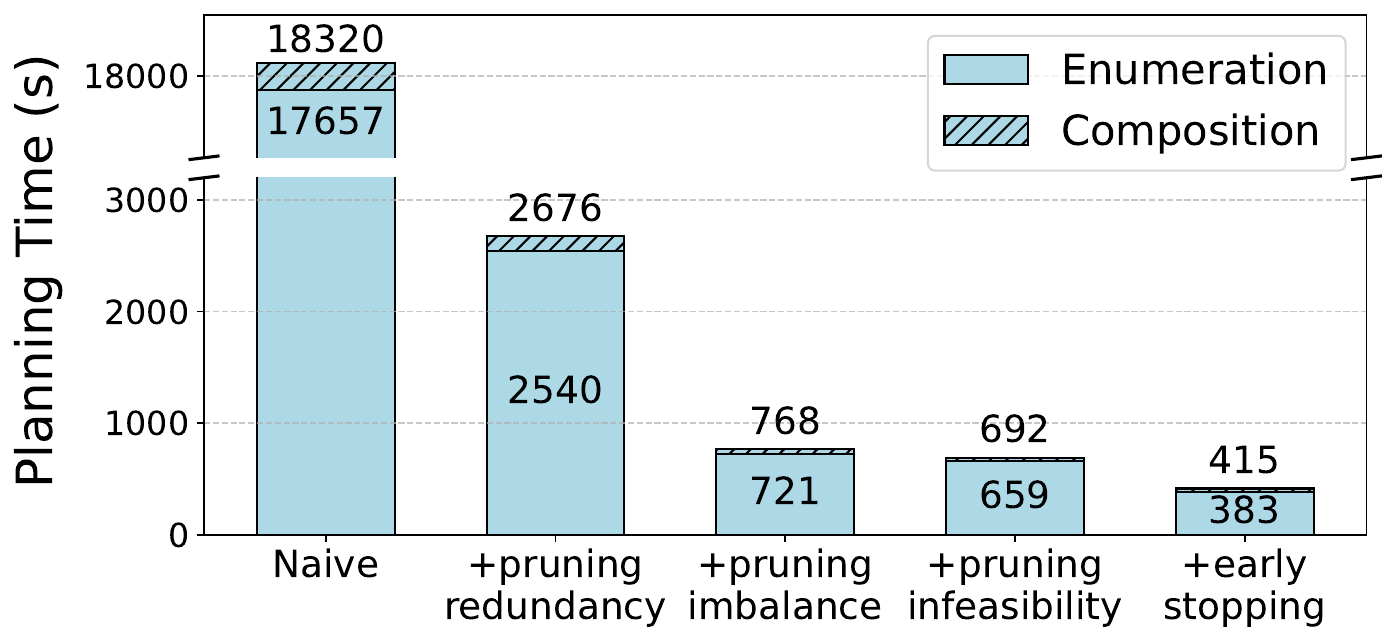}
  \caption{Breakdown of the impact of pruning policies on end-to-end planning time \textnormal{(GPT-3 $6.7$B model, MBS=1, cluster~\textsf{C1-compute-het} with 8$\times$V100 and 8$\times$T4 GPUs)}}
  \label{fig:hetml_search_time_breakdown}
\end{figure}

We study \sys's scalability due to the effectiveness of its pruning policies. Our ablation study shows a breakdown of the impact of the pruning policies on end-to-end planning time for \sys-Aceso in our experiment from \S\ref{sec:eval:het-skew}.

As \F\ref{fig:hetml_search_time_breakdown} shows, naively enumerating all \msi plans takes $5.1$\unit{h}. Applying all pruning policies speeds up the search by $27\times$---an order of magnitude. Pruning redundant \msi pairs with the same number of transformer layers has the biggest impact: a reduction of $7\times$. Pruning unbalanced \msi pairs before enumeration also contributes significantly, reducing planning time by a further $4\times$. Pruning infeasible and incompatible \msi pairs adds a planning time reduction only by $1.1\times$, down to $12$\unit{min}. This is expected, because infeasible/incompatible pairs are themselves unbalanced, therefore already filtered out. Overall, the enumeration time of \msi plans dominates compared to the composition time. Most importantly, the proposed pruning policies do not lead to a performance degradation, because the composed plans remain the same, even after a significant reduction in \msi plans during pipeline composition.

We also include an Aceso-specific~\cite{liu2024aceso} \emph{early-stopping} strategy. Aceso normally employs a large fixed time budget for its iterative search of the best parallelization plan, terminating early when the budget is exceeded. In \sys, Aceso only parallelizes \msi pairs, which makes it possible to adjust this search budget dynamically. Since model slices range from small to large in terms of transformer layers, we scale this budget with the size of the model slice relative to the full LLM. This yields a further $1.7\times$ reduction in search time, with no observed impact on plan quality.

The ablation study shows that pruning policies significantly reduce planning time without affecting plan quality.

\subsection{Scalability of parallelization planning}
\label{sec:eval:search}

\begin{figure}[t]
  \centering
  \includegraphics[width=\columnwidth]{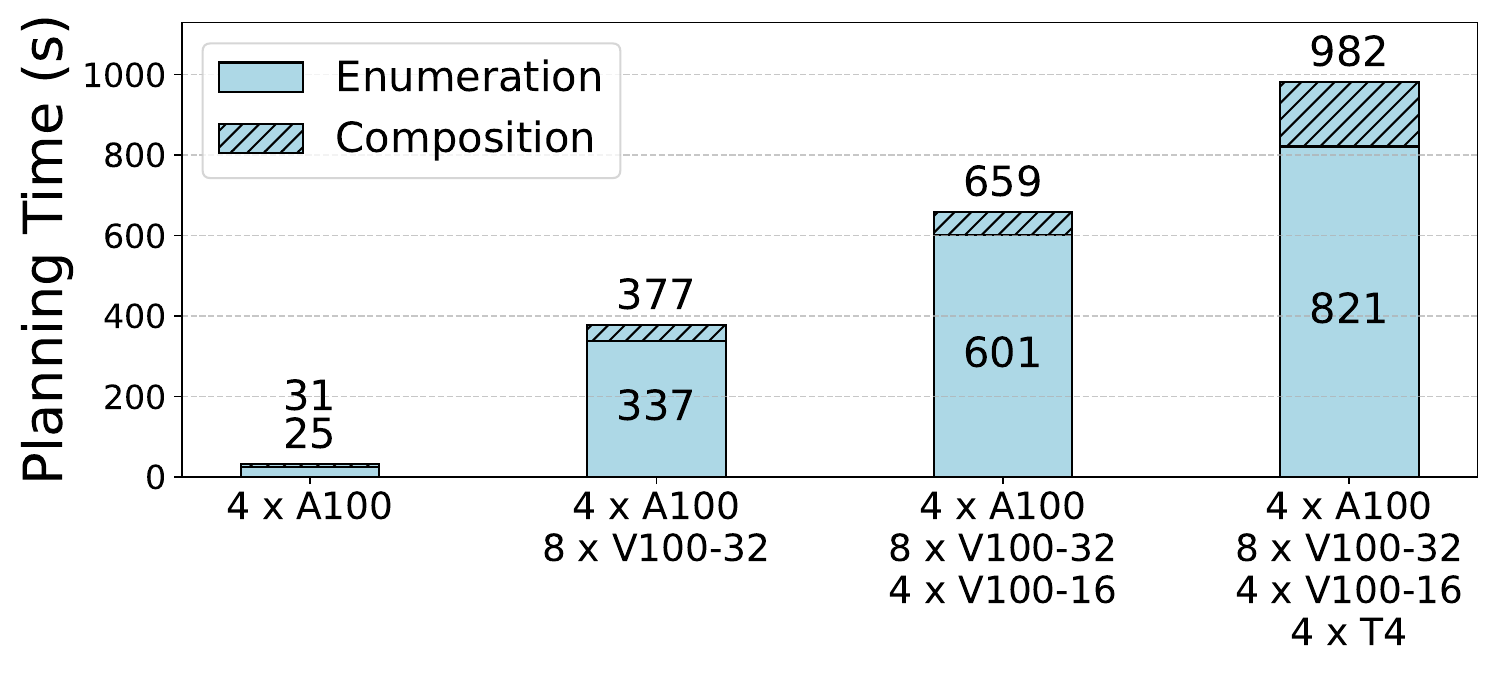}
  \caption{Planning time when scaling GPU cluster heterogeneity \textnormal{(40-layer GPT-3 $13$B model, MBS=1, cluster~\textsf{C3-full-het})}}
  \label{fig:hetml_search_scaling}
\end{figure}

We also explore how \sys{}'s end-to-end planning time is influenced by the degree of heterogeneity, \ie the number of GPU islands, as the cluster size increases. We consider parallelization planning for GPT-3 $13$B with $1$--$4$~islands. The cluster setup is the same as in~\S\ref{sec:eval:heterogeneity}, and we deploy \sys-Aceso with the full set of pruning policies.

\F\ref{fig:hetml_search_scaling} shows the planning time for each configuration. With 1~island, using \sys-Aceso is equivalent to using Aceso by itself, which takes $31$\unit{s}. \sys only requests model-slice/island plans for a single model slice, \ie one that contains the full model. With 2~islands, the planning time increases to $377$\unit{s}---a $3.8\times$ increase compared to the chosen parallelization budget given to Aceso. Adding 3~islands increases the time to $659$\unit{s}, which is another increase of $1.7\times$. Finally, with 4~islands, the planning time increases by $1.5\times$.

Similar to \S\ref{sec:eval:pruning}, enumeration outweighs the pipeline composition for all island counts, but introduced heterogeneity comes at a cost. The planning time, despite the enumeration of model-slice/island plans, is around $16.4$\unit{min}, even for a 40-layer model , and a cluster of 20~GPUs and 4~GPU islands. This is reasonable compared to the substantially longer training times. It shows that \sys's approach of hiding heterogeneity and pruning policies allows it to scale to larger LLM training scenarios with significant heterogeneity, while still finding high-quality heterogeneous plans.

\section{Conclusion}
\label{sec:conclusion}

We described \sys, a system for LLM training that hides heterogeneity from parallelization planning. \sys discovers homogeneous GPU islands through profiling, and decomposes the LLM into model slices. Each model slice is then parallelized on a GPU island by utilizing existing homogenenous LLM parallelizers. \sys composes these \msi pairs through load-balanced pipeline parallelism. To find the best pipeline, \sys enumerates potential \msi pairs and composes them into a pipeline. Through a parallel search and heterogeneity-guided pruning, \sys accelerates this search by an order of magnitude. \sys achieves substantial speedups over heterogeneous LLM parallelizers on heterogeneous GPU clusters, while matching homogeneous LLM parallelizers on homogeneous clusters. 



\bibliographystyle{ACM-Reference-Format}
\bibliography{main}

\end{document}